\documentclass[12pt, letter]{article}

\usepackage[margin=1in]{geometry} 

\usepackage{times} 
\usepackage[utf8]{inputenc}
\usepackage{microtype} 
\usepackage{setspace}

\usepackage{amsmath}
\usepackage{amsfonts}
\usepackage{amssymb}

\usepackage{graphicx}      
\usepackage{booktabs}      
\usepackage{caption}       
\usepackage{subcaption}    
\usepackage{floatrow}      
\usepackage{float}         
\usepackage{siunitx}
\usepackage{makecell}
\usepackage{threeparttable}

\usepackage[round]{natbib} 

\usepackage[colorlinks=true,
    linkcolor=black,
    urlcolor=black,
    citecolor=black]{hyperref}

\usepackage{array}

\title{\textbf{At-Risk Transformation for U.S. Recession Prediction} \\$~$}

\author{
  Rahul Billakanti \\ \textit{Wayzata High School}
  \and
  Minchul Shin \\ \textit{FRB Philadelphia}\thanks{We thank Todd Clark, Frank Diebold, Domenico Giannone, Laura Liu, Benjamin Malin, Massimiliano Marcellino, Kenwin Maung, and Jonathan Wright for valuable comments. The views expressed herein are those of the authors and do not necessarily reflect the views of the Federal Reserve Bank of Philadelphia, or the Federal Reserve System. Emails: rahulsai.billakanti11@gmail.com, minchul.shin@phil.frb.org.}
  \\ $~$
  \\ $~$
}

\date{December 2025} 

\begin{document}

\maketitle 

\doublespacing

\begin{abstract}
We propose a simple binarization of predictors—an ``at-risk'' transformation—as an alternative to the standard practice of using continuous, standardized variables in recession forecasting models. By converting predictors into indicators of unusually weak states based on a thresholding rule estimated from training data, we demonstrate their ability to capture the discrete nature of rare events such as U.S. recessions. Using a large panel of monthly U.S. macroeconomic and financial data, we show that binarized predictors consistently improve out-of-sample forecasting performance, often making linear models competitive with flexible machine learning methods, and that the gains are particularly pronounced around the onset of recessions. 
\end{abstract}

\vspace{1cm} 

\newpage 


\section{Introduction}
\label{sec:intro}
The accurate and timely forecasting of U.S. recessions remains a central challenge in macroeconomics, with direct implications for policymakers, investors, and households. The recent literature has largely advanced along two main fronts: (i) identifying informative predictors through variable selection, screening, or factor-based aggregation of large datasets \citep{EstrellaMishkin1998, StockWatson1993_RecessionProcedure, Chen2011}, and (ii) exploiting flexible non-parametric and machine learning methods to uncover nonlinearities that may improve recession prediction \citep{Qi2001, ng2014boosting, vrontos2021modeling}.

Despite these advances, a common feature of this literature is the treatment of predictors. Most studies, whether focusing on variable selection, combining large sets of predictors, or employing nonlinear and non-parametric methods, typically rely on raw data inputs. Beyond standard procedures such as stationarization and standardization, they rarely apply additional variable transformations. Drawing on the papers cited throughout this study, as well as a targeted review of articles from the \textit{International Journal of Forecasting}, we find that existing work on recession nowcasting and forecasting generally avoids nonlinear transformations of predictors.\footnote{We review all papers on model-based recession forecasting published in the \textit{International Journal of Forecasting} as of August 2025. Of course, if we broaden from binary recession prediction to continuous targets like output growth or inflation, there is work using nonlinear predictor transformations \citep[e.g.,][]{gouletcoulombe2021MDTM, Daniele2025}. In addition, there are applications beyond recession prediction in which nonlinear transformations of predictors are explicitly adopted in logistic regression. For example, in a recent paper, \citet{liu2025binary} analyze binary outcomes with heavy-tailed covariates and show that their semiparametric tail objective is asymptotically equivalent to a logistic regression on tail observations using the logarithm of the extreme predictor. See Appendix \ref{ssec:additional_lit} for a more detailed analysis.} Instead, these approaches feed the raw variables directly into the forecasting model, leaving any nonlinearities to be captured by the parametric or nonparametric modeling framework itself.

In this paper, we take a different perspective. We propose that a crucial nonlinearity can be embedded directly into the predictors themselves before any model is estimated. To do this, we apply what we call the ``at-risk'' transformation: a simple binarization of predictors prior to their inclusion in otherwise standard forecasting models. Specifically, each predictor is converted into an indicator variable that equals one when the series enters an unusually weak state relative to its historical distribution. While such a transformation may appear to discard valuable information, we argue that it is well-suited for predicting rare events, such as U.S. recessions, where the relevant signal often lies in whether indicators cross into unusually adverse territory.

Using a large panel of monthly U.S. macroeconomic and financial data, we show that binarized predictors consistently outperform standard continuous predictors in out-of-sample recession forecasting. The improvement is robust across horizons, model classes, and aggregation methods. In linear settings, logistic regressions with binarized inputs yield higher discriminatory power than their continuous counterparts. In nonlinear settings, additional tree-based methods provide little incremental benefit once binarization is applied, indicating that much of the relevant nonlinearity is already captured by the transformation itself. We further show that combining binarized variables through methods such as principal components improves predictive performance. 

This paper contributes to the rare-event forecasting literature and practice \citep[e.g.,][]{LahiriYang2013} by showing that substantial predictive gains can be achieved by reconsidering the representation of predictors. We introduce a simple, data-dependent ``at-risk'' transformation that converts predictors into indicators of unusually weak states. This transformation consistently improves forecasting performance across linear and nonlinear models, and across multiple horizons. Because it is easy to implement and computationally inexpensive, the approach offers a practical benchmark for both academic research and applied forecasting of rare events such as recessions or defaults. Our proposal is also closely related to several strands of the existing literature--including early-warning rules, diffusion indexes, business cycle dating methods, and tree-based approaches--which we review in Section \ref{ssec:related_literature}.

The remainder of the paper is organized as follows. 
Section~\ref{sec:methodology} introduces the proposed at-risk transformation and associated prediction methods, and relates them to the existing literature. 
Section~\ref{sec:oos} describes the out-of-sample forecasting design and evaluation metrics. 
Section~\ref{sec:results} presents the main empirical findings.  
Section~\ref{sec:understanding} investigates the sources of these gains by comparing probability forecasts and variable importance across models. 
Finally, Section~\ref{sec:parsimony} concludes by verifying that the findings hold in parsimonious predictor sets. We provide additional robustness checks in the appendix.

\section{Predicting U.S. Recessions with the At-Risk Transformation}
\label{sec:methodology}

We begin with a standard model for forecasting U.S. recessions,
\[
P(y_{t+h} = 1) = f(X_{t}' \theta),
\]
where $y_{t+h}$ is a recession indicator that equals one if the U.S. economy is in a recession, as defined by the NBER, at time $t+h$. $X_{t}$ is an $N \times 1$ vector of predictors that are useful for forecasting U.S. recessions. The function $f(\cdot)$ maps the predictors to the recession probability. A popular choice for $f(\cdot)$ leads to logistic or Probit regression. In our setup, time is measured monthly, and we consider $h=3,6,12$, corresponding to 3-month, 6-month, and 12-month-ahead forecasts.

This general framework encompasses most prediction models that have appeared in the literature. There are numerous studies on what to include in $X_{t}$. For example, some researchers find that the yield curve is a strong predictor of recessions \citep[e.g.,][]{estrella1996yield, wright2006yield, rudebusch2009}. More recently, researchers have found that including a large set of economic indicators can improve predictive performance, either by applying dimensionality reduction techniques to $X_{t}$ before estimating a logistic regressions or by employing regularization methods in conjunction with logistic regression.

Another important choice is the functional form of $f(\cdot)$ and the linear index $X_{t}'\theta$. Some researchers find that non-parametric approaches, such as Random Forests or gradient-boosted trees, can perform better than parametric approaches like logistic regression \citep{ng2014boosting, Dopke2017, vrontos2021modeling}. Others find that deep learning models such as LSTM and GRU neural networks show promising results \citep{Qi2001,chung2024insideblackboxneural}. Overall, the literature suggests that incorporating nonlinearity can yield sizable predictive gains relative to a plain-vanilla logistic regression model where $X_{t}$ enters linearly.

\subsection{At-Risk Transformation}

Economic downturns are often preceded by persistent weakness in certain indicators. Rather than using the entire range of variation in a predictor, it may be more informative to focus on whether that predictor is in an unusually weak state relative to its own historical behavior. To formalize this idea, we transform each stationary series $x_{it}$ into a new binary variable $z_{it}$, which we call the \textit{at-risk} transformation.


For each variable $x_{it}$, the \textit{at-risk} transformation is defined as
\begin{equation} \label{eq:at-risk}
z_{it} =  
\mathbf{1}\!\left\{\, s_i \bar{x}^{h_g}_{it} \le Q_{i,h_g}(\tau_g) \,\right\},
\end{equation}
where $s_{i}\in\{+1,-1\}$ indicates the cyclical orientation of the variable, with $s_{i}=+1$ for pro-cyclical variables and $s_{i}=-1$ for counter-cyclical variables; $\bar{x}_{it}^{h_{g}}$ is the $h_{g}$-month moving average of the original series $x_{it}$,
\[
\bar{x}_{it}^{h_{g}} = \frac{1}{h_{g}} \sum_{s=0}^{h_{g}-1} x_{i,t-s},
\]
and $Q_{i,h_g}(\tau_g)$ denotes the $\tau_g$-quantile of the historical distribution of $s_i \bar{x}^{h_g}_{it}$.

This transformation converts each original series $x_{it}$ into a binary indicator $z_{it}$ that equals one if the historical moving average of $x_{it}$ falls below a specified threshold, given by the $\tau_{g}$-th quantile of its historical distribution. The two parameters $\tau_{g}$ and $h_{g}$ are the key tuning parameters of the transformation. In the next section, we discuss how we select these tuning parameters and provide practical guidance. In our empirical analysis, the classification of each variable as pro-cyclical or counter-cyclical is predetermined, with the full list provided in Appendix \ref{ssec:CCvars}. This classification is based primarily on economic theory and, when needed, supported by historical correlations from the training data. 

The core idea is that recessions are more closely associated with extreme values--the tail behavior--of certain predictors rather than with their full range of variation. By focusing on these tail observations, the \textit{at-risk} transformation aims to improve predictive performance relative to using the raw series $x_{it}$ directly. Whether this approach improves forecasting accuracy is ultimately an empirical question; our results show that this specific nonlinear transformation provides meaningful gains in predicting U.S. recessions.

\subsection{Aggregating At-Risk Signals}

Once the \textit{at-risk} transformation is applied, the prediction model becomes 
\[
P(y_{t+h} = 1) = f(Z_{t}' \theta),
\]
where $Z_{t} = [z_{1t}, z_{2t}, ..., z_{Nt}]'$ collects the $N$ transformed binary indicators. This formulation links the set of binary predictors directly to the recession indicator. Interpreting each $z_{it}$ as the output of an individual ``recession signal'' model, the single index $Z_{t}' \theta$ represents an aggregated signal obtained by combining these individual signals. The function $f(\cdot)$ then maps the aggregated signal into a predicted recession probability.

As we demonstrate in the empirical section the ``at-risk'' transformation converts a large panel of continuous macroeconomic predictors into a binary state matrix, $Z_t$. While this matrix may contain predictive information, its high dimensionality presents a significant modeling challenge. Using the full, disaggregated matrix directly in a predictive model carries a high risk of overfitting and can lead to unstable parameter estimates. Therefore, another central empirical question we will explore in this study is how to best aggregate and model this information to maximize out-of-sample forecasting performance. To this end, we consider several aggregation strategies.

\paragraph{Aggregation Strategies.} Each aggregation strategy reduces $Z_{t}$ to a lower-dimensional representation $W_{t}$, where $\dim(W_{t}) \ll \dim(Z_{t})$:

\begin{quote}
\begin{description}
    \item[Disaggregated (Baseline)] As a direct benchmark, our baseline model uses the full $Z_{t}$ matrix of binarized predictors without further aggregation.
    
    \item[Simple Average] The predictor is the cross-sectional mean of the disaggregated binarized predictors:
\[
W_t = \frac{1}{n} \sum_{i=1}^{n} z_{it}
\]
similar to the ``counting rule'' widely used in constructing diffusion indexes, with \citet{NBERc0729} providing the classic example.

    \item[Unsupervised Aggregation (PCA)] The predictors are the first $K$ principal components of $Z_t$:
    \[
    W_t = V_k' \cdot Z_t,
    \]
    where $V_K$ is the $N \times K$ matrix whose columns are the eigenvectors corresponding to the $K$ largest eigenvalues of the sample covariance matrix of $Z_t$. In our study, we set $K=8$, following \citet{mccracken2016fred}.
    
\end{description}
\end{quote}

\paragraph{Main prediction models.} For each aggregation approach, we consider two predictive model classes:

\begin{quote}
\begin{description}
    \item[Logistic Regression] We specify $f(\cdot)$ as a logistic link applied to the linear index $W_{t}' \theta$. In the baseline case using the full $Z_{t}$, we employ $\ell_{2}$ penalization (Ridge) to mitigate overfitting from high dimensionality.
    \item[Gradient Boosting] We also consider a non-linear tree-based model (XGBoost). While the at-risk transformation already introduces nonlinearity, tree-based methods can capture complex interactions among predictors that may further improve predictive performance.
\end{description}
\end{quote} 




\subsection{Binarized Predictors and Related Literature} \label{ssec:related_literature}

The idea of transforming economic indicators into threshold-based signals has appeared in several strands of the literature, most prominently in the early-warning system framework. A well-known example in recession prediction is the \textit{negative spread rule}, where an inverted yield curve (spread $<0$) is interpreted as a warning signal of an impending recession \citep{laurent1988, lahiri2023b}. Similarly, the \textit{Sahm rule} identifies downturns when the three-month moving average of the unemployment rate rises by 0.50 percentage points or more relative to the minimum of the three-month averages from the previous 12 months, a specific threshold \citep{sahm2019}. 

The closest antecedent to our proposed method is \cite{KeilisBorok2000}, who adapted a pattern recognition algorithm from earthquake prediction to forecast U.S. recessions. Their approach transformed six economic indicators into binary signals and issued alarms whenever a sufficient number crossed pre-specified thresholds. They exemplified that such binary transformations contain predictive information beyond that captured by linear models. Our contribution builds on this insight by extending the use of binarized variables to a high-dimensional setting. Rather than relying on a small, fixed set of indicators and heuristic aggregation rules, we apply the \textit{at-risk} transformation to a broad panel of macroeconomic predictors and explore multiple data-driven aggregation methods to combine these signals into recession probabilities.

Averaging a large number of binary signals has also been adopted in analyzing business cycles. For example, the classic work of \citet{NBERc0729} introduced a diffusion index constructed as a simple average of binary indicators, a method recently revisited by \citet{mathy2025could}, who, in their context, analyzed whether these diffusion indexes could predict recessions like the Great Depression. 

More generally, the so-called ``date-and-average'' approach to business cycle dating builds on the same principle: individual series are first classified into contraction or expansion phases, after which these binary classifications are aggregated to determine the overall chronology \citep[see, for example ][]{burns1946measuring, harding2006synchronization, stock2014estimating, crump2020reading}. The aggregation is typically implemented by exploiting the clustering of turning points across series. This logic naturally motivates our approach of transforming continuous predictors into binary contraction/expansion signals and generating forecasts from their historical association with the NBER recession indicator. 

Our method is also related to tree-based approaches, since our binarized variables can be interpreted as restricted decision stumps with data-dependent cutoffs. While we show in our results that a simple linear aggregation of binarized predictors often outperforms standard tree-based methods, we also find that tree-based models can offer advantages in specific contexts, particularly for improving forecast performance when using aggregated factors. 

\section{Out-of-Sample Forecast Evaluation}
\label{sec:oos}

\subsection{Data}
Our empirical analysis relies on the FRED-MD monthly database, a standard dataset for macroeconomic forecasting and business cycle research developed by \citet{mccracken2016fred}. The sample spans January 1960 to December 2024 and includes 126 (reduced to 122 after exclusions) time series covering key sectors of the economy, such as output and income, housing activity, and financial markets. All series are transformed following the procedures in \citet{mccracken2016fred}. The target variable ($y_t$) is the NBER recession indicator, \texttt{USRECM}, which equals 1 in recession months and 0 otherwise.

\subsection{Model Specification and Tuning Parameters}
\label{ssec:params}
\paragraph{Model specification.}
Our baseline model includes $Z_{t}$ and its lags as predictors in both the logistic regression and XGBoost specifications. We define the set of included lags, $\mathcal{L}$, as $\mathcal{L} = \{3, 6, 12\}$, which includes short-, medium-, and long-term lags. We use a constant set of lags to not force any assumptions about optimal horizon-dependent lags—this allows the model to select the best features across them. The same specification is applied to models with $X_{t}$ as well as to those using the aggregated versions of $Z_{t}$ and $X_{t}$.

\paragraph{Tuning parameters.}
The at-risk transformation contains two tuning parameters, $\tau_{g}$ and $h_{g}$. The quantile level $\tau_{g}$ determines the threshold at which we convert continuous data into binary indicators. This quantile is computed from the empirical distribution of the moving average $\bar{x}^{h_{g}}_{it}$, where $h_{g}$ is the window size. To avoid look-ahead bias, we set $\tau_{g}$ in the forecasting performance evaluation as follows: we take the median of the median individual quantiles during recession periods, based on the initial training sample (January 1960 to December 1989). This procedure leverages historical information on when the warning signal would have been activated. We then freeze this value for the remainder of the evaluation sample to ensure that no future information is used. We present a formal algorithm to select $\tau_{g}$ in Appendix \ref{sssec: tauselect}. Our default choice for the window size is $h_{g}=1$ (meaning no smoothing). Larger values allow the model to incorporate lagged information into the predictors themselves, which could, in principle, improve forecasts. In practice, however, we find that including lags of $Z_{t}$ while keeping $h_{g}=1$ delivers strong predictive performance.

These rules are deliberately simple yet practical, providing a natural baseline. In the robustness exercises reported in the Appendix, we show that there is potential for further predictive gains by fine-tuning these parameters. For example, $\tau_{g}$ could be allowed to vary across different categories of predictors or the predictors themselves (Appendix \ref{ssec:varythresholds}). In addition, $h_{g}$ can be greater than one to incorporate more lagged information (Appendix \ref{ssec:varysmooth}).  While such refinements may improve forecast accuracy, we keep the baseline specification simple to highlight how well even these parsimonious choices perform.

\paragraph{Computation.}
When the logistic regression includes many variables, we employ $\ell_{2}$ regularization, with the penalty strength selected by time-series cross-validation on the initial training sample and then held fixed for the remainder of the evaluation period to avoid look-ahead bias (see Appendix \ref{sssec:selectreg} for full procedure). For PCA, we fix the number of factors for $X_{t}$ at eight, following \citet{mccracken2016fred}, and use the same number for $Z_{t}$. All computations are implemented in Python: logistic regression models are estimated by maximum likelihood with an $\ell_{2}$ penalty (via {\small \texttt{scikit-learn}}), while nonlinear benchmarks are estimated using gradient boosting with $\ell_{2}$ regularization on leaf weights (via {\small \texttt{xgboost}}). Unless otherwise noted, XGBoost is run with its default hyperparameters, which yield stable and competitive performance in our application.

\subsection{Evaluation Methods}
\label{subsec:evaluation}
To ensure our results reflect true predictive ability, our entire evaluation is conducted in a strict out-of-sample context. We use a recursive forecasting design, with an initial training window from 1960 to 1989. We use this to produce the first forecast for January 1990. The training data is then expanded by one month for each subsequent forecast, and all model parameters and aggregation weights are re-estimated at every step to prevent any lookahead bias. 

Throughout the text, we assess forecast quality using two primary metrics for a binary classification problem:

\paragraph{Precision–Recall Area Under Curve (PR AUC).} 
Our model produces predicted recession probabilities, $\hat{p}_{t}$, which can be converted into point forecasts using a decision threshold $\delta \in [0,1]$:
\[
\hat{y}_{t}(\delta) = \mathbf{1}\{\hat{p}_{t} \geq \delta\}.
\]
For each $\delta$, we compute recall (true positive rate) and precision (positive predictive value) over the evaluation sample:
\[
R(\delta) = \frac{TP(\delta)}{TP(\delta) + FN(\delta)}, 
\qquad 
P(\delta) = \frac{TP(\delta)}{TP(\delta) + FP(\delta)},
\]
where $TP$, $FN$, and $FP$ denote true positives, false negatives, and false positives, respectively. 
Varying $\delta$ traces out the precision–recall curve. 
The precision–recall area under the curve is then defined as
\[
\text{PR AUC} = \int_{0}^{1} P(\delta) \, \frac{dR(\delta)}{d\delta} \, d\delta
= \int P(R) \, dR,
\]
which integrates precision with respect to recall as the decision threshold varies. 

PR AUC is particularly well-suited for imbalanced datasets such as recession forecasting, since it emphasizes performance on the rare positive class. Unlike ROC AUC, which measures performance in terms of false positive rate, PR AUC places direct weight on precision. This means it penalizes false alarms much more strongly, an important advantage in imbalanced settings such as recession forecasting, where non-recession periods outnumber recession months. For a detailed discussion of this issue, as well as comparisons between ROC AUC and PR AUC in the context of recession forecasting and related studies, see \citet{lahiri2023} and the references therein. See Appendix \ref{ssec:roc} for an ROC AUC description and scores from preliminary comparisons.

The baseline PR AUC equals the unconditional probability of a recession,
\[
\frac{\#\{\text{recession months}\}}{\#\{\text{total months}\}}
= \frac{36}{420} \approx 0.086,
\]
meaning that random guessing would achieve a PR AUC of 0.086. This value provides a natural lower bound: any model should exceed this benchmark, and higher PR AUC values are strictly better, with 1 representing a perfect model.

\paragraph{Brier Score.} 
The Brier Score is a proper scoring rule that measures the mean squared error of probabilistic forecasts:
\[
\text{BS} = \frac{1}{n} \sum_{t=1}^{n} (\hat{p}_{t} - y_t)^2,
\]
where $\hat{p}_{t}$ is the predicted probability and $y_t$ is the realized outcome (0 for expansion, 1 for recession). 
Like the mean squared error of point forecasts, the Brier Score is a function of terms that reflect reliability (calibration, analogous to bias) and resolution (discrimination, analogous to variance) \citep{murphy1973new, DieboldRudebusch1989}. 
A random guess of $\hat{p}_{t} = 0.5$ yields a Brier Score of $0.25$, which serves as a useful upper bound. 
Smaller values indicate better accuracy and calibration of the probabilistic forecasts, with $0$ representing a perfect model.

\section{Main Empirical Findings}
\label{sec:results}

This section provides the central empirical evidence for the at-risk transformation. We first benchmark the predictive value of $Z_t$ against continuous-input and factor-model alternatives. We then assess whether alternative aggregation schemes or nonlinear models further improve performance. Finally, we use forecast encompassing tests to evaluate whether competing approaches add information beyond our proposed framework.

\subsection{The Predictive Value of the ``At-Risk'' Transformation}
\label{ssec:atriskpower}
We begin our empirical analysis by evaluating the performance of the at-risk transformation within a traditional linear framework by comparing a Logistic Regression model with an $\ell_{2}$ (Ridge) regularization penalty trained on the disaggregated $Z_t$ matrix (our baseline model) against models trained on the full $X_t$ predictor set, as well as a standard factor model benchmark using Principal Component Analysis. The comprehensive out-of-sample results for the 3-, 6-, and 12-month forecast horizons are presented in \autoref{tab:zt_vs_xt}. For each model-input combination, we report the PR AUC and Brier Score. Other metrics are reported in the appendix.

\begin{table}[t!]
\centering
\caption{Out-of-Sample Performance of the At-Risk Transformation ($Z_t$) vs. Benchmarks}
\label{tab:zt_vs_xt}
\begin{tabular}{l ccc ccc}
\toprule
& \multicolumn{3}{c}{\textbf{PR AUC}} & \multicolumn{3}{c}{\textbf{Brier Score}} \\
\cmidrule(lr){2-4} \cmidrule(lr){5-7}
\textbf{Model Configuration} & \textbf{$h=3$} & \textbf{$h=6$} & \textbf{$h=12$} & \textbf{$h=3$} & \textbf{$h=6$} & \textbf{$h=12$} \\
\midrule

\textbf{Proposed ($Z_t$, Logit-$\ell_{2}$)} & 
\textbf{0.718} & 
0.370 & 
\textbf{0.398} & 
\textbf{0.049} & 
\textbf{0.082} & 
\textbf{0.087} \\
\addlinespace

\midrule
\textbf{Alternative Specifications} &&&&& \\
$X_t$, Logit-$\ell_{2}$ & 
0.501 & 
0.286 & 
0.170 & 
0.069 & 
0.096 & 
0.121 \\

PCA of $X_t$, Logit-$\ell_{2}$ & 
0.552 & 
\textbf{0.408} & 
0.314 & 
0.064 & 
0.083 & 
0.108 \\

$X_t$, XGBoost & 
0.584 & 
0.338 & 
0.351 & 
0.062 & 
0.085 & 
0.098 \\
\bottomrule
\end{tabular}
\footnotesize
\begin{flushleft}
\textit{Note:} All models use the full predictor set and include contemporaneous values plus 3-, 6-, and 12-month lags. The 'Proposed' model is a logistic regression with an $\ell_2$ penalty trained on the binarized 'at-risk' state matrix ($Z_t$). Benchmarks are trained on the standard continuous data matrix ($X_t$). Bold values indicate the best-performing model for each metric and horizon.
\end{flushleft}
\end{table}

The out-of-sample results show that our proposed 'at-risk' transformation ($Z_t$) provides a distinct advantage at all horizons. The improvement in probabilistic accuracy is particularly notable, as evidenced by the consistently lower Brier Scores of the proposed model. The proposed model also shows a notable increase in discriminatory power, especially on $h=3$.

Consistent with previous studies, extracting the common component from $X_t$ prior to its inclusion in the predictive model appears to enhance overall predictive power. In addition, we also find that allowing $X_{t}$ to enter the predictive model nonlinearly improves performance (e.g., $X_{t}$ with XGBoost) relative to its linear counterpart (e.g., $X_{t}$ with Logit-$\ell_{2}$). However, our proposed model outperforms both these alternative and traditional specifications. This suggests that our ``at-risk'' transformation appears to capture a highly relevant form of nonlinearity for recession prediction.

The performance at the medium-term ($h=6$) horizon reveals a more complex dynamic. The PR AUC for our proposed model is 0.370, slightly lower than its performance at the 12-month horizon (0.398). This non-monotonic pattern across horizons has been observed in other studies \citep{vrontos2021modeling}. Importantly, we find that this concavity disappears under alternative specifications, for example, when using PCA on $Z_{t}$, suggesting that the shape of the performance curve is sensitive to how the at-risk signals are aggregated. Nonetheless, even at the $h=6$ horizon, our proposed approach continues to deliver better-calibrated probabilities and higher overall accuracy than traditional alternatives.

Taken together, the evidence in \autoref{tab:zt_vs_xt} provides strong support for our central hypothesis. The ``at-risk'' transformation, which converts a large panel of continuous data into a matrix of binary state indicators, creates a powerful and robust feature set for recession forecasting. Our proposed framework, even when implemented with a linear specification and simple regularization, proves superior to, or highly competitive with, benchmarks that include standard factor models and more complex non-linear classifiers. Having established the fundamental value of the $Z_t$ matrix, we next turn to methods for refining this signal through aggregation.

\subsection{Evaluation of Advanced Modeling Strategies}
\label{ssec:modeling_strategies_results}
Having established the inherent value of the $Z_t$ matrix, we now investigate whether its signal can be refined and enhanced. In this section, we conduct a comparison of the different aggregation and modeling strategies outlined in our methodology to identify the optimal modeling approach for the at-risk feature set.

\autoref{tab:aggregation_comparison} presents the out-of-sample performance of our three primary aggregation strategies: the disaggregated baseline, a simple average, and an unsupervised PCA approach. To examine the interaction between aggregation and model complexity, we evaluate each strategy using both Logistic Regression with $\ell_{2}$ regularization and an XGBoost model across all three forecasting horizons. As in the previous section, we present the PR AUC and Brier Scores of aggregation method under Logistic Regression and XGBoost.

\begin{table}[t!]
  \centering
  \caption{Out-of-Sample Performance of Alternative Aggregation Strategies for $Z_t$}
  \label{tab:aggregation_comparison}
  \resizebox{\textwidth}{!}{
  \begin{tabular}{l l ccc c ccc}
    \toprule
    & & \multicolumn{3}{c}{\textbf{PR AUC}} & & \multicolumn{3}{c}{\textbf{Brier Score}} \\
    \cmidrule(r){3-5} \cmidrule(l){7-9}
    \textbf{Aggregation Method} & \textbf{Model} & \textit{$h=3$} & \textit{$h=6$} & \textit{$h=12$} & & \textit{$h=3$} & \textit{$h=6$} & \textit{$h=12$} \\
    \midrule
    Disaggregated (Baseline) & Logit-$\ell_{2}$ & \textbf{0.718} & 0.370 & 0.398 & & \textbf{0.049} & 0.082 & 0.087 \\
                                & XGBoost & 0.583 & 0.300 & 0.269 & & 0.065 & 0.089 & 0.099 \\
    \addlinespace
    Simple Average & Logit-$\ell_{2}$ & 0.541 & 0.365 & 0.190 & & 0.151 & 0.188 & 0.229 \\
                     & XGBoost & 0.466 & 0.257 & 0.106 & & 0.115 & 0.133 & 0.141 \\
    \addlinespace
    PCA on $Z_t$ & Logit-$\ell_{2}$ & 0.688 & 0.528 &\textbf{0.404} & & 0.062 & 0.081 & 0.106 \\
                       & XGBoost & 0.686 & \textbf{0.567} & 0.315 & & 0.052 & \textbf{0.063} & \textbf{0.086} \\
    \bottomrule
    \addlinespace
  \end{tabular}
  }
  \parbox{\textwidth}{\footnotesize \textit{Note:} This table compares the performance of different methods for modeling the $Z_t$ matrix. The "Disaggregated" row corresponds to the baseline model from Section \ref{ssec:atriskpower}. Bold values indicate the best performance for each metric and horizon. }
\end{table}

Taken together, the results in \autoref{tab:aggregation_comparison} shed light on how different aggregation strategies interact with model complexity in shaping predictive performance. Several key patterns emerge.

First, the Logit-$\ell_{2}$ models exhibit significantly higher discriminatory power, often without a meaningful loss in Brier Score, especially for short horizons. This underscores that simple linear classifiers remain highly effective when paired with our at-risk transformation.

Second, as noted earlier for the disaggregated baseline, XGBoost does not consistently improve performance. In fact, adding additional nonlinearity often reduces predictive accuracy. This contrasts with the results in \autoref{tab:zt_vs_xt}, where applying XGBoost to the raw $X_{t}$ dataset substantially boosted its performance. The difference suggests that the “at-risk” transformation effectively linearizes the prediction problem with respect to the recession indicator. By embedding the essential nonlinearities into the features themselves, it allows a parsimonious and robust linear model to outperform more complex alternatives.

Third, the simple average aggregation, which is analogous to a diffusion index or ``counting rule,'' performs poorly. By treating all indicators equally, it effectively washes out valuable heterogeneity across predictors. This result underscores the importance of allowing weights to differ across signals rather than merely counting the number of indicators currently at risk.

Lastly, PCA on $Z_{t}$ performs particularly well at the medium- and long-horizon forecasts ($h=6,12$). In terms of PR AUC, it often matches or exceeds the disaggregated baseline, while also producing calibrated probabilities with competitive Brier Scores. For short horizons ($h=3$), the disaggregated benchmark still performs best, but PCA remains close. Notably, PCA on $Z_{t}$ delivers the lowest Brier Score at the 6-month horizon when paired with XGBoost, suggesting that it can refine predictive accuracy in some settings. From a practical, real-time forecasting perspective, PCA on $Z_t$ offers a significant advantage. Macroeconomic data are released asynchronously, leaving a ``ragged edge'' at the end of the sample. The PCA framework provides a natural and established method for handling this missing data. This makes PCA on $Z_{t}$ an appealing real-time forecasting tool, complementing the disaggregated specification’s strong short-horizon performance.

\subsection{Forecast Encompassing Test}
From the metrics themselves, we can conclude that our approaches ($Z_t$ and $Z_t$ with PCA) are superior to their continuous-input counterparts. However, it is important to formally test whether they provide statistically significant information beyond that contained in standard benchmarks. To this end, we implement a forecast encompassing test. This type of test evaluates whether one forecast contains all the relevant information in another and is therefore more efficient \citep{granger1973}. Early applications include \citet{fair1990comparing}, and later work extends the framework to probability forecasts by \cite{clements2010}. Our implementation follows the same spirit but differs in detail: rather than projecting the target on forecast probabilities, we estimate a probit regression of the binary recession indicator on the log-odds of the competing probability forecasts.

More specifically, we estimate a probit model where the dependent variable, $y_t$, is the NBER recession indicator at time $t$. The regressors are the out-of-sample predicted probabilities from our proposed model and from a competing benchmark. To ensure the regressors are unbounded, we apply the log-odds (logit) transformation, $L(\hat{p_t}) = \log\left(\dfrac{\hat{p_t}}{1-\hat{p_t}}\right)$, to each probability series. The estimated model is specified as
\begin{equation}\label{eq:encompassingreg}
P(y_t=1 \mid \hat{p}_{t, A}, \hat{p}_{t, B}) = \Phi(\beta_0+\beta_A \cdot L(\hat{p}_{t, A}) + \beta_B \cdot L(\hat{p}_{t, B})),
\end{equation}
where $\hat{p}_{t, A}$ is the forecast from our proposed model (A), $\hat{p}_{t, B}$ is the forecast from the benchmark model (B), and $\Phi(\cdot)$ denotes the standard normal cumulative distribution function. Since our goal is to test whether model $A$ (proposed) encompasses model $B$ (benchmark), we focus on the significance of the coefficient $\beta_{B}$. If $\beta_{B}=0$, the benchmark forecast contributes no additional predictive power once we condition on the forecasts from the proposed model.

\begin{table}[t!]
    \centering
    \caption{Forecast Encompassing Test Results}
    \label{tab:encompassing_tests}
    \sisetup{
        table-format=-1.2,
        add-integer-zero=true,
        table-space-text-post=***
    }

    \begin{tabular}{@{} l l S S S S @{}}
        \toprule
        & & \multicolumn{2}{c}{$\beta_A$} & \multicolumn{2}{c}{$\beta_B$} \\
        \cmidrule(lr){3-4} \cmidrule(lr){5-6}
        \textbf{Proposed Model (A)} & \textbf{Benchmark Model (B)} & {Coeff.} & {p-value} & {Coeff.} & {$p$-value} \\
        \midrule
        \addlinespace
        
        $\underline{h=3}$ &&&&&\\
        $Z_t$ (Logit-$\ell_{2}$) & $X_t$ (Logit-$\ell_{2}$)         & \num{0.630}\textsuperscript{***} & 0.000 &  0.003 & 0.958 \\
        $Z_t$ (Logit-$\ell_{2}$) & $X_t$ (XGBoost)        & \num{0.592}\textsuperscript{***} & 0.000 &  0.026 & 0.689 \\
        PCA on $Z_t$ (Logit-$\ell_{2}$) & PCA on $X_t$ (Logit-$\ell_{2}$) & \num{0.548}\textsuperscript{***} & 0.000 & 0.040 & 0.405 \\
        \midrule
        \addlinespace
        
        $\underline{h=6}$ &&&&&\\
        $Z_t$ (Logit-$\ell_{2}$) & $X_t$ (Logit-$\ell_{2}$)         & \num{0.331}\textsuperscript{***}  & 0.000 &  0.049 & 0.444 \\
        $Z_t$ (Logit-$\ell_{2}$) & $X_t$ (XGBoost)        & \num{0.298}\textsuperscript{***} & 0.000 &  0.049 & 0.288 \\
        PCA on $Z_t$ (Logit-$\ell_{2}$) & PCA on $X_t$ (Logit-$\ell_{2}$) & \num{0.587}\textsuperscript{***} & 0.000 &  0.020 & 0.730 \\
        \midrule
        \addlinespace
        
        $\underline{h=12}$ &&&&&\\
        $Z_t$ (Logit-$\ell_{2}$) & $X_t$ (Logit-$\ell_{2}$)         & \num{0.595}\textsuperscript{***} & 0.000 &  -0.097 & 0.405 \\
        $Z_t$ (Logit-$\ell_{2}$) & $X_t$ (XGBoost)        & \num{0.350}\textsuperscript{***} & 0.005 &  \num{0.077}\textsuperscript{*} & 0.069 \\
        PCA on $Z_t$ (Logit-$\ell_{2}$) & PCA on $X_t$ (Logit-$\ell_{2}$) & \num{0.746}\textsuperscript{***} & 0.000 &  0.308 & 0.243 \\
        
        \bottomrule
        \addlinespace
    \end{tabular}
    
    \captionsetup{justification=justified, singlelinecheck=false}
    \parbox{\textwidth}{\footnotesize 
        \textit{Note:} The table reports the estimated coefficients ($\beta$) and corresponding $p$-values from the probit encompassing regression specified in \autoref{eq:encompassingreg}. $Z_t$ refers to the disaggregated ``at-risk'' matrix, and $X_t$ refers to the full FRED-MD dataset. Significance at the 10\%, 5\%, and 1\% levels is denoted by *, **, and ***, respectively.}
\end{table}


The results of the forecast encompassing tests, presented in \autoref{tab:encompassing_tests}, provide strong and consistent statistical evidence for the superiority of our proposed approaches. In most cases, the coefficient on the benchmark model ($\beta_B$) is statistically indistinguishable from zero, indicating that benchmark forecasts contribute no incremental information. There are a few instances where $\beta_B$ is significantly different from zero, but even then the magnitude of $\beta_A$ is much larger than that of $\beta_B$, implying that forecasts from our proposed model dominate those from benchmark models based on $X_t$. Taken together, these results reinforce the findings from the previous sections that the at-risk transformation delivers superior predictive content relative to traditional specifications.

\section{Understanding Drivers of Forecasting Performance}
\label{sec:understanding}

To better understand the relative strengths of our baseline “at-risk” transformation model and the continuous-predictor benchmark, this section examines both the forecasts they generate and the economic drivers underlying those forecasts. We first analyze the out-of-sample probabilities to assess how timely and decisive each model is in signaling recessions. We then turn to the composition of forecast importance across variables and sectors to evaluate whether differences in performance can be traced to different economic underpinnings.

\subsection{Out-of-Sample Probabilities}
\begin{figure}[!t!]
    \centering
    \caption{Out-of-Sample Recession Probabilities of Disaggregated $Z_t$}
    \label{fig:main_probability_plots}

    \begin{subfigure}{0.88\textwidth}
        \includegraphics[width=\textwidth]{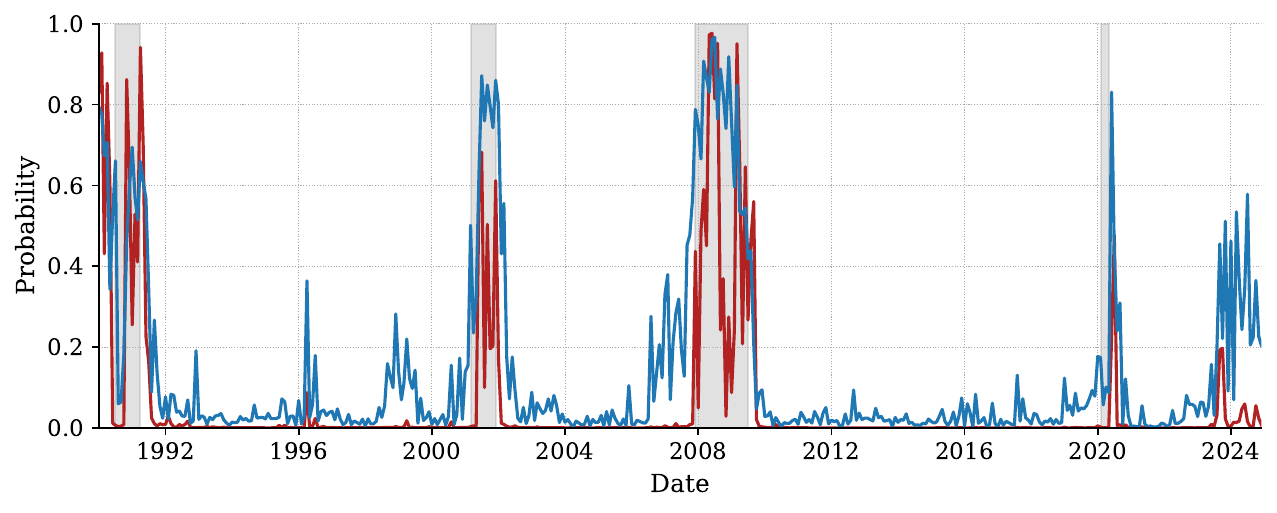}
        \caption{3-Month-Ahead Forecast ($h=3$)}
        \label{fig:h3_prob_plot}
    \end{subfigure}
    
    \begin{subfigure}{0.88\textwidth}
        \includegraphics[width=\textwidth]{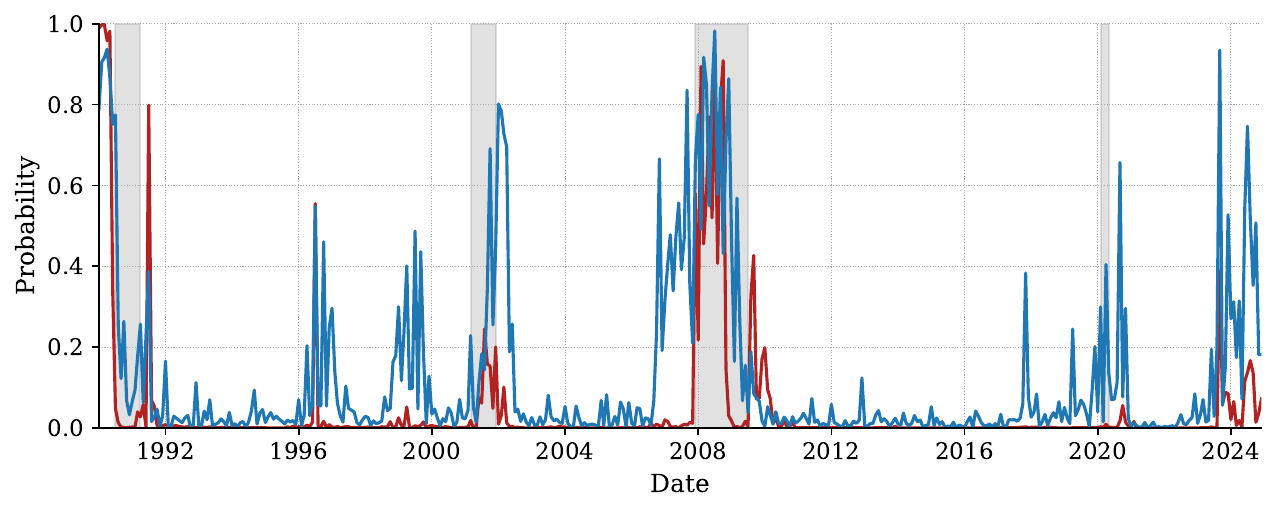}
        \caption{6-Month-Ahead Forecast ($h=6$)}
        \label{fig:h6_prob_plot}
    \end{subfigure}

    \begin{subfigure}{0.88\textwidth}
        \includegraphics[width=\textwidth]{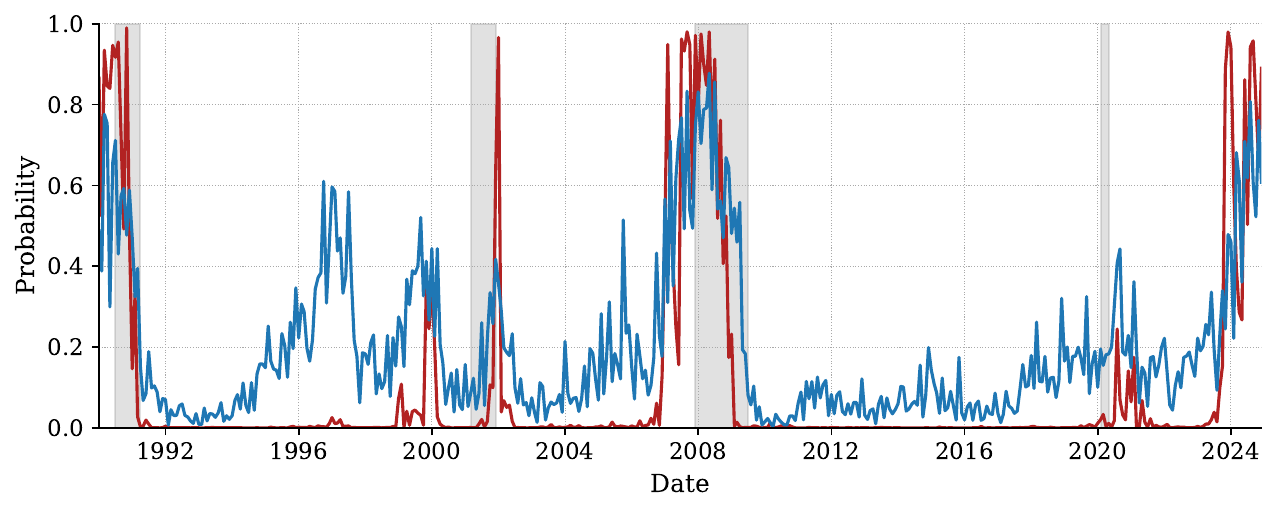}
        \caption{12-Month-Ahead Forecast ($h=12$)}
        \label{fig:h12_prob_plot}
    \end{subfigure}
    \parbox{\textwidth}{\footnotesize \textit{Note:} The figure plots the out-of-sample monthly recession probabilities from our primary $Z_t$ (Logit-L2) model (blue line) and the Full $X_t$ (XGBoost) model (red line). Shaded vertical bars indicate official NBER recession periods.}
\end{figure}

First, we start by examining the probability forecasts generated by the models. To visualize this, \autoref{fig:main_probability_plots} presents the out-of-sample monthly recession probabilities from our standard ``at-risk'' transformation—the Disaggregated $Z_t$ (Logit-L2) model and the Full $X_t$ (XGBoost) model. The figure displays the forecasts for each of the three horizons ($h=3, 6, 12$ months), allowing for a direct comparison of how our binary variables perform against their continuous counterparts.

The plots reveal several aspects of our proposed model and its relation to its continuous benchmark. First, the probabilities of $Z_t$ rise much more rapidly than the probabilities of $X_t$ on every horizon. The model's probabilities generally spike in recessions such as 1990, 2001, and 2008, indicating that the binarized transformation more effectively captures the onset and dynamics of recessions than continuous variables.

Another observation is the cautiousness of the XGBoost model trained on the Full FRED-MD dataset. Across $h=3,6,12$, its probabilities stay very dormant in expansionary periods, even when there is economic turmoil that did not lead to an official recession. But this indecisiveness by the model is the cause for its subpar performance. It tends to remain very cautious and raises its probabilities only during the clearest signals of a downturn. For example, on $h=6$, it only raises probabilities to slightly above 0.2 during the 2001 recession (which was known for being relatively mild), and on $h=12$, it completely misses the recession and raises its probabilities afterward.

On the other hand, our approach offers a better point of performance on this trade-off. In signs of clear economic expansion (such as the 2010s), it keeps its probabilities very low on all horizons, leading to a better Brier Score. It is also more sensitive, making it faster to react and keeping more sustained probabilities during the recession, resulting in it having superior discrimination (PR AUC). Thus, our framework offers a very good balance between sensitivity and calibration, something that the continuous predictor model struggles with.

\begin{figure}[!t!]
    \centering
    \includegraphics[width=0.95\textwidth]{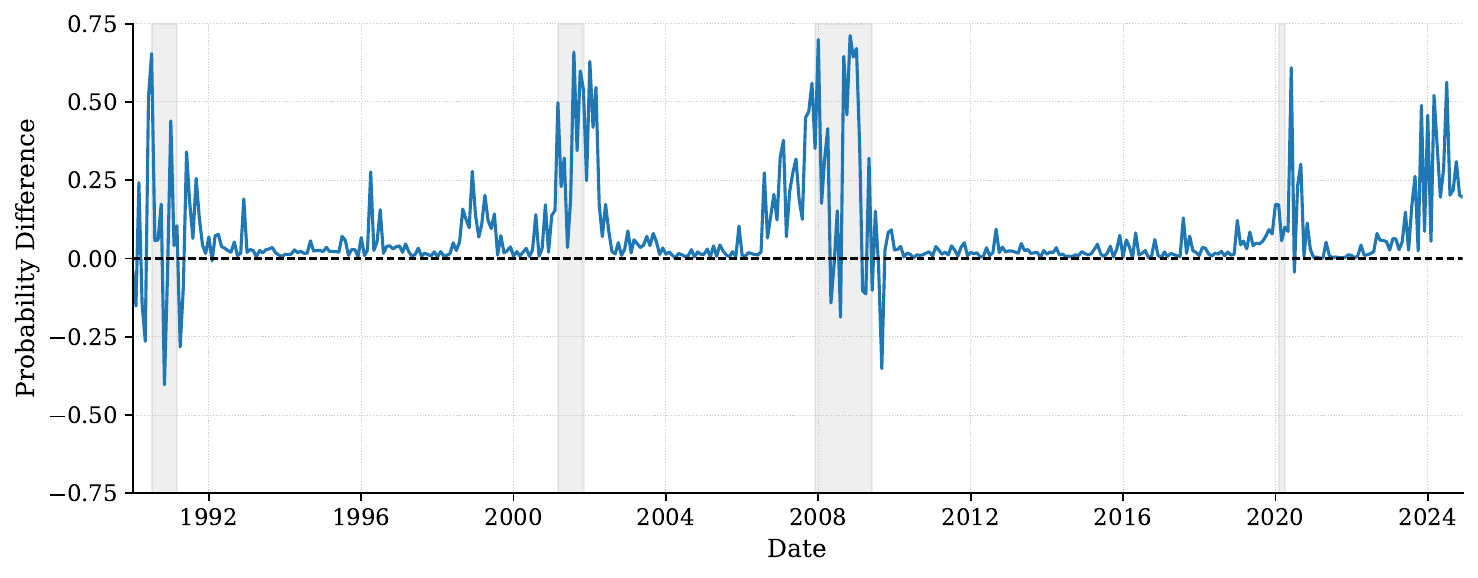}
    \caption{Forecast Disagreement Between Proposed and Benchmark Models ($h=3$)}
    \label{fig:disagreement_h3}
    \parbox{\textwidth}{\footnotesize \textit{Note:} The figure plots the difference between the out-of-sample recession probability from our primary $Z_t$ (Logit-L2) model and the $X_t$ (XGBoost) benchmark for the 3-month-ahead forecast. Positive values indicate that our proposed model assigned a higher probability of recession. Shaded vertical bars indicate official NBER recession periods.}
\end{figure}

To further demonstrate this, we begin with a visual analysis of the forecast disagreement between our $Z_t$ (Logit-L2) model our benchmark, $X_t$ (XGBoost), for the 3-month horizon. \autoref{fig:disagreement_h3} plots the time series of the difference between their out-of-sample recession probabilities. Positive values indicate periods where our "at-risk" model assigned a higher probability of recession than the XGBoost benchmark, while negative values indicate the reverse.

The plot reveals a revealing pattern about the nature of using binary and continuous predictors. In the periods immediately preceding the NBER-dated recessions of 1990, 2001, 2008, and 2020, the disagreement series consistently spikes into positive territory. This indicates that our ``at-risk'' transformation provides earlier and more decisive warning signals than the benchmark. Additionally, while somewhat variable, the disagreement series tends to reside in positive values during recessionary periods.

During expansionary periods, the disagreement series generally hovers around 0, indicating that there is not a substantial difference between the probabilities. However, it does lie on the positive side, confirming our finding from \autoref{fig:main_probability_plots} that the $X_t$ and XGBoost configuration is well-calibrated during expansions. To further illustrate this phenomenon, we decompose the Brier Score (Mean Squared Error) across recessionary and expansionary periods for each model, found in \autoref{tab:sse_decomposition_h3}.

\begin{table}[t!]
    \centering
    \caption{Decomposition of Mean Squared Error (MSE) for $h=3$ Forecasts}
    \label{tab:sse_decomposition_h3}
    \begin{tabular}{lcc}
        \toprule
        \textbf{Model Configuration} & \textbf{Proposed Model} & \textbf{Benchmark Model} \\
        & ($Z_t$, Logit-L2) & ($X_t$, XGBoost) \\
        \midrule
        MSE during Recessions & 0.222 & 0.513 \\
        MSE during Expansions & 0.031 & 0.015 \\
        \midrule
        \textbf{MSE on Full Sample} & \textbf{0.049} & \textbf{0.062} \\
        \bottomrule
        \addlinespace
    \end{tabular}
    
    \parbox{\textwidth}{\footnotesize \textit{Note:} The table reports the Mean Squared Errors (MSE/Brier Score) for the 3-month-ahead out-of-sample forecasts, decomposed into NBER-dated recession and expansion periods. A lower value indicates a better forecast calibration.}
    
\end{table}
As noted in \autoref{fig:disagreement_h3}, the XGBoost model trained on $X_t$ is slightly more calibrated on expansions, though the difference is small. By contrast, the Brier Score for the $X_t$ and XGBoost model is significantly higher than that of our approach on strictly recessionary periods, illustrating that our approach provides a much better trade-off in terms of sensitivity and calibration.

\subsection{Economic Drivers of Forecast Performance}
\label{ssec:drivers}
The preceding sections established the superior out-of-sample performance of models using the ``at-risk'' transformation. We now turn to an analysis of the economic drivers of this performance gain.

We first examine which specific features contribute to the forecasts made by the considered $Z_t$ and $X_t$ models to understand the key driver of increased predictive performance due to the ``at-risk'' transformation. The top 10 predictors with the highest average coefficients/importances (summed across all lags) are reported in \autoref{tab:feature_importance}. We take $h=3$ as our chosen example for this analysis.

\begin{table}[t!]
  \centering
  \caption{Top 10 Most Important Predictors by Framework ($h=3$)}
  \label{tab:feature_importance}
  \sisetup{table-align-text-post=false}
  \begin{tabular}{l S[table-format=1.3] l S[table-format=2.3]}
    \toprule
    \multicolumn{2}{c}{\textbf{``At-Risk'' ($Z_t$) + Logit-L2}} & \multicolumn{2}{c}{\textbf{Continuous ($X_t$) + XGBoost}} \\
    \cmidrule(r){1-2} \cmidrule(l){3-4}
    Variable & {Importance} & Variable & {Importance} \\
    \midrule
    3 mo-FF spread & 0.704 & 5 yr-FF spread & 93.733 \\
    1 yr-FF spread & 0.694 & 10 yr-FF spread & 71.358 \\
    6 mo-FF spread & 0.653 & 1 yr-FF spread & 70.255 \\
    M2 (real) & 0.649 & 6 mo-FF spread & 68.517 \\
    5 yr-FF spread & 0.623 & 3 mo-FF spread & 26.535 \\
    CP-FF spread & 0.622 & Emp: total & 26.478 \\
    10 yr-FF spread & 0.594 & Avg hrs: mfg & 17.736 \\
    S\&P div yield & 0.575 & Aaa-FF spread & 17.389 \\
    Aaa-FF spread & 0.520 & Starts: nonfarm & 17.373 \\
    S\&P 500 & 0.491 & Emp: mfg & 14.949 \\
    \bottomrule
    \addlinespace
  \end{tabular}
  \parbox{\textwidth}{\footnotesize \textit{Note:} Importance scores are aggregated across all relevant lags for each base variable. The importances of the Logit-$\ell_2$ model are the average absolute coefficient. The importances of the XGBoost model are the average `Gain' metric. The absolute scales of the two metrics are not directly comparable; the analysis focuses on the relative rankings and composition of the predictor sets.}
\end{table}


The results reveal fundamental similarities and differences in the models’ learned strategies. Both models assign a central role to interest rates and term spreads in forecasting recessions, consistent with the classic findings of \citet{estrella1996yield}. However, the distribution of importance diverges sharply: in the $X_t$ + XGBoost model, feature importance falls rapidly from about 70 for spread variables to roughly 27 for the leading labor market indicator, indicating that the model concentrates weight on a narrow set of predictors. By contrast, the $Z_t$ + Logit-L2 model shows a gradual tapering of importance across predictors, with monetary aggregates (e.g., M2) and stock market indicators retaining nontrivial influence. Although both models incorporate $\ell_2$ penalization, XGBoost produces a much sharper hierarchy, placing disproportionate emphasis on a small subset of variables, whereas our baseline model distributes weight more evenly across the predictor set.

\begin{table}[t!]
  \centering
  \caption{Sectoral Contribution to Forecasts in Pre-Recession Periods}
  \label{tab:sector_contribution}
  \sisetup{table-align-text-post=false}

\begin{tabular}{l *{4}{>{\centering\arraybackslash}p{1.2cm}}}
    \toprule
     \multicolumn{5}{c}{\textbf{Contribution to Forecast (\%) in Year Before Recession}} \\
    \midrule
    \multicolumn{5}{c}{\textbf{`At-Risk' ($Z_t$) + Logit-$\ell_{2}$ Model}} \\
    \midrule
    Economic Sector $\phantom{abcdeee}$ & {1990} & {2001} & {2008} & {2020} \\
    \midrule
    Output \& Income & 10.8 & 8.8 & 9.7 & 10.1 \\
    Labor Market & 26.9 & 28.0 & 27.2 & 26.5 \\
    Housing & 7.1 & 8.0 & 8.1 & 7.8 \\
    Consumption \& Orders & 4.5 & 4.3 & 4.5 & 5.1 \\
    Money \& Credit & 10.5 & 10.6 & 10.1 & 10.8 \\
    Interest Rates \& Spreads & 27.1 & 25.9 & 25.6 & 24.3 \\
    Prices & 8.1 & 9.3 & 9.4 & 9.2 \\
    Stock Market & 5.1 & 5.2 & 5.4 & 6.2 \\
    \midrule
    \multicolumn{5}{c}{\textbf{Continuous ($X_t$) + XGBoost Model}} \\
    \midrule
    Economic Sector $\phantom{abcdeee}$ & {1990} & {2001} & {2008} & {2020} \\
    \midrule
    Output \& Income & 3.4 & 2.4 & 3.8 & 6.3 \\
    Labor Market & 19.5 & 20.2 & 30.4 & 20.2 \\
    Housing & 2.5 & 11.9 & 13.2 & 13.7 \\
    Consumption \& Orders & 0.7 & 1.7 & 2.8 & 1.0 \\
    Money \& Credit & 0.8 & 5.3 & 1.6 & 2.1 \\
    Interest Rates \& Spreads & 66.1 & 52.6 & 43.8 & 53.8 \\
    Prices & 0.8 & 0.6 & 1.2 & 1.0 \\
    Stock Market & 6.3 & 5.4 & 3.2 & 1.9 \\
    \bottomrule
    \addlinespace
  \end{tabular}
  \parbox{\textwidth}{\footnotesize \textit{Notes:} The table reports the percentage of total feature importance attributable to each economic sector during the 12-month window immediately preceding the NBER-dated peak of each recession. Values are calculated from the out-of-sample feature importances of the `$Z_t$, Logit-L2' and `$X_t$, XGBoost' models.}
\end{table}

We further illustrate this pattern in \autoref{tab:sector_contribution}, which reports the average contribution of each economic sector—following the classification of \citet{mccracken2016fred}—to forecasts during the 12 months preceding each NBER recession peak. The table reinforces our earlier finding: the $X_t$ model relies disproportionately on interest rates and spreads, often assigning them more than half of total importance, while giving only marginal weight to other sectors. By contrast, the “at-risk” transformation yields a more balanced structure. Interest rates and labor market indicators emerge as the two leading contributors, each carrying a similar share of importance, with other categories—such as prices, money and credit, and output and income—also playing more meaningful roles. Such diversification highlights the mechanism through which the at-risk transformation improves predictive performance compared with the continuous specification.

Another important observation from \autoref{tab:sector_contribution} is the stability of sectoral contributions across the four recessions in our baseline model. In contrast, the $X_t$ + XGBoost model exhibits considerable fluctuations in its allocation of importance. For example, the contribution of Interest Rates \& Spreads and Stock Market variables declines across the first three recessions, while the influence of Labor Market and Housing variables rises. Given the stronger performance of our baseline model, these shifts in the $X_t$ model are more plausibly attributable to overfitting than to genuine time-variation in the underlying predictive relationships. Binarization appears to make the forecasting framework more robust by preventing it from responding excessively to short-term variation in $X_t$ and instead focusing on the tail behavior that carries the strongest recession signals.

\section{Parsimonious Modeling with At-Risk Transformation}
\label{sec:parsimony}
Our analysis of the full, high-dimensional models has revealed two key findings: first, that our ``at-risk'' transformation is the superior out-of-sample performer, and second, that extracting factors from a high-dimensional $Z_t$ rather than $X_t$ is a better modeling choice. However, this motivates a crucial final question: is the ``at-risk'' transformation's benefit solely a phenomenon of ``big data,'' or does it represent a fundamental improvement applicable to simpler, more traditional forecasting exercises?

To answer this, we conduct the same out-of-sample analysis used for our main results, but in a low-dimensional environment. We consider two predictor sets. The first contains only a single variable—the 10-year Treasury–Fed Funds spread. The second contains ten canonical predictors emphasized in the recession forecasting literature, which we refer to as the parsimonious model.\footnote{To construct a fair and robust comparison, we use indicators not derived directly from our feature importances. Instead, we follow the academic literature to select variables that mimic a traditional forecasting exercise: two spreads (10-year Treasury–Fed Funds spread and Baa–Fed Funds spread), three labor market measures (total employment, unemployment insurance claims, and the unemployment rate), and one representative variable from each of the other major categories: industrial production (real activity), nonfarm housing starts (housing), retail sales (consumption and orders), the S\&P 500 index (stock market), and real M2 (money and credit).} For each of these reduced predictor sets, we evaluate the performance of a Logit-$\ell_2$ model applied to the at-risk transformation ($Z_t$) against an XGBoost model applied to the corresponding continuous variables ($X_t$).

\begin{table}[t!]
  \centering
  \caption{Out-of-Sample Performance: Univariate and Parsimonious Models}
  \label{tab:subset}
  \sisetup{
    table-align-text-post=false,
    table-number-alignment = center
  }
  \begin{tabular}{l *{3}{S[table-format=1.3]} *{3}{S[table-format=1.3]}}
    \toprule
    & \multicolumn{3}{c}{\textbf{PR AUC}} & \multicolumn{3}{c}{\textbf{Brier Score}} \\
    \cmidrule(lr){2-4} \cmidrule(lr){5-7}
    Model Specification & {$h=3$} & {$h=6$} & {$h=12$} & {$h=3$} & {$h=6$} & {$h=12$} \\
    \midrule
        \multicolumn{7}{l}{\textit{Full Feature Space}} \\
    \quad Continuous ($X_t$) + XGBoost & 0.584 & 0.338 & 0.351 & 0.062 & 0.085 & 0.098 \\
    \quad ``At-Risk'' ($Z_t$) + Logit-$\ell_{2}$ & \textbf{0.718} & \textbf{0.370} & \textbf{0.398} & \textbf{0.049} & \textbf{0.082} & \textbf{0.087} \\
    \midrule
    \multicolumn{7}{l}{\textit{Univariate Models (Term Spread Only)}} \\
    \quad Continuous ($X_t$) + XGBoost  & \textbf{0.367} & 0.242 & 0.255 & \textbf{ 0.092} & \textbf{0.111} & \textbf{0.109} \\
    \quad ``At-Risk'' ($Z_t$) + Logit  & 0.264 & \textbf{0.334} & \textbf{0.384} & 0.130 & 0.124 & 0.121 \\
    \midrule
    \multicolumn{7}{l}{\textit{Parsimonious Models (10 Core Indicators)}} \\
    \quad Continuous ($X_t$) + XGBoost  & 0.529 & 0.314 & 0.343 & 0.072 & 0.093 & \textbf{0.090} \\
    \quad ``At-Risk'' ($Z_t$) + Logit-$\ell_2$  & \textbf{0.721} & \textbf{0.485} & \textbf{0.451} & \textbf{0.057} & \textbf{0.088} & 0.117 \\
    \bottomrule
    \addlinespace
  \end{tabular}
    \parbox{\textwidth}{\footnotesize 
  \textit{Notes:} The ``Full Feature Space'' corresponds to the entire FRED-MD panel used in earlier sections. 
  The ``Univariate'' specification uses only the 10-year Treasury–Fed Funds spread. 
  The ``Parsimonious'' model uses ten core indicators grounded in the literature: two spreads (10-year Treasury–Fed Funds spread, Baa–Fed Funds spread), three labor market measures (total employment, unemployment insurance claims, unemployment rate), industrial production (real activity), nonfarm housing starts (housing), retail sales (consumption and orders), the S\&P 500 index (stock market), and real M2 (money and credit).
  }
\end{table} 

The results are presented in \autoref{tab:subset}. For reference, the top panel reproduces the full feature space results from the previous section, where we showed that the at-risk transformation outperforms the continuous $X_{t}$ representation even when the latter is paired with a nonlinear XGBoost model. The second panel, which examines the univariate performance of both approaches with the 10-year Fed Funds term spread, reveals a nuanced result. The continuous representation is superior in terms of overall calibration at all horizons, but the binarized term spread exhibits significantly better discriminatory power (PR AUC) at $h=6, 12$. Additionally, the ``at-risk'' state is effective at longer horizons with long-lead indicators. However, we find that when relying on a univariate feature space, the effects of binarization vary by indicator and horizon.

The last panel, however, reveals the true power of our approach. In a realistic multivariate setting, one that mirrors traditional recession forecasting practice, the at-risk transformation ($Z_{t}$) consistently outperforms the continuous specification ($X_{t}$) across nearly every horizon and metric, with especially large gains in discriminatory power (PR AUC) at medium and long horizons. This confirms our main finding that the ``at-risk'' transformation is a robust and effective way to improve recession predictability. 


Although the parsimonious model tested here is not a fine-tuned specification and is built on a generic set of indicators chosen for their economic intuition, a researcher does not know a priori which indicators to include in a prediction model. Therefore, one should not interpret the performance presented in this table as evidence that this set can dominate other aggregation methods in practice. The point of this exercise is to demonstrate that the strong and consistent outperformance of the $Z_{t}$ framework in this setting provides compelling evidence of its value beyond high-dimensional applications (see Appendix \ref{ssec:otherparsimony} for additional combinations).

While the transformation's effect on any single variable can be complex, its primary strength is its ability to provide diverse, yet unified, signals, allowing a simple model to learn from a chorus of evidence (discrete 1s and 0s). For recession forecasting problems relying on a handful of key indicators, the ``at-risk'' transformation offers a more effective method for representing the predictive information contained in these key indicators.

\section{Conclusion}
The evidence presented in this paper suggests that a simple binarization of predictors—the ``at-risk'' transformation—is a powerful tool for recession forecasting. Building on the foundational insight of \citet{KeilisBorok2000}, who first applied this idea to a small set of indicators in a simpler setup, we demonstrate the effectiveness of a similar approach in a modern, high-dimensional forecasting environment. Our recursive out-of-sample analysis shows that models using these binary features are not only highly competitive but often superior to benchmarks that use standard continuous data, including machine learning methods like XGBoost. We also find that performance improves significantly when extracting PCA factors from the binarized representation of continuous variables.

Looking ahead, several directions can broaden the scope of our findings. First, as with any forecasting model, the out-of-sample period, while spanning over three decades and multiple business cycles, is ultimately finite. The definitive test of the framework's robustness will be its continued performance in real-time. Second, it would be interesting to test the idea in other countries, using comparable datasets \citep[e.g.,][for the UK]{gouletcoulombe2021}, and in other contexts such as quantile regression \citep[e.g.,][]{adrian2019vulnerable}. Lastly, it would be fruitful to formalize the data-generating settings under which binarized “at-risk” predictors and their aggregations dominate continuous-input models. A natural case is a common extreme-shock mixture, where mean shifts are negligible but the frequency of synchronized tail realizations spikes in pre-recession states—making counts or principal components of standardized $Z_t$ near-sufficient while linear models on $X_t$ are misspecified. This also connects to nonlinear PCA for binary data (e.g., logistic PCA), providing a principled counterpart to our pragmatic PCA benchmark.




\bibliographystyle{apalike} 
\bibliography{references} 

\newpage

\appendix

\begin{center}
{\huge{ \textbf{Appendix}}}    
\end{center}

\section{Method Details}
\subsection{Data}
For this study, we rely on the FRED-MD macroeconomic dataset, containing monthly data for 126 time series in eight categories. Similar to the authors' exercise, we remove ACOGNO (New Orders for Consumer Goods), TWEXAFEGSMTHx (Nominal Advanced Foreign Economies U.S. Dollar Index), UMCSENTx (Consumer Sentiment Index), and OILPRICEx (Crude Oil, spliced WTI and Cushing) because either they are highly irregular after transformation or a significant amount of data is missing \citep{mccracken2016fred}, resulting in 122 macroeconomic predictors. 

A crucial part of time-series analysis is the stationarity of the predictor variables. To achieve this, we follow the standard procedure for the FRED-MD dataset by applying the specific transformations recommended by \citet{mccracken2016fred}. 
\subsection{Counter-cyclical variables}
\label{ssec:CCvars}
We use a mix of judgment and data-driven analysis to determine which stationarized series are counter-cyclical. Since the stationarized series $x_t$ behaves differently from the original series from FRED-MD, we compute its correlation with the NBER recession indicator on the in-sample period. If the correlation is strongly negative ($<-0.10$), the series is categorized as a counter-cyclical series. The remaining series are classified as pro-cyclical.

The variables classified as counter-cyclical are as follows: unemployment rate (U: all), mean duration of unemployment rate (U: mean duration), civilians unemployed \textless \ 5 weeks (U \textless \ 5 wks), civilians unemployed 5-14 weeks (U 5-14 wks), civilians unemployed 15+ weeks (U 15+ wks), civilians unemployed 15-26 weeks (U 15-26 wks), civilians unemployed 27+ weeks (U 27+ wks), initial claims (UI claims), inventories to sales ratio (M\&T invent/sales), and CBOE volatility index.

\subsection{Hyperparameter Selection}
\subsubsection{Selecting $\tau_g$}
\label{sssec: tauselect}
For the global threshold $\tau_g$ employed in the main text, we determine it once from the initial training sample ($t=1,...,T$) to prevent look-ahead bias. The procedure is as follows:
\begin{enumerate}
    \item For each predictor $i$ and for each recession month $t \in S_{rec}$, we compute the empirical quantile level, $\tau_{it}$, of the observation $\bar{x}_{it}^{h_{g}}$ relative to the full training-sample distribution, $F_{i, T}$.
    \[ \tau_{it} = F_{i, T}(\bar{x}_{it}^{h_{g}}) \]

    \item For each predictor $i$, we then calculate the median of these recession-time quantile levels. This value, $\tau_i^*$, represents the typical quantile level for that specific predictor during historical recessions.
    \[ \tau_i^* = \text{median}(\{\tau_{it} \mid t \in S_{rec}\}) \]

    \item Finally, the global threshold $\tau_g$ is set to the median of these predictor-specific values. This approach balances the signals across all $N$ predictors in the dataset.
    \[ \tau_g = \text{median}(\{\tau_i^* \mid i = 1, ..., N\}) \]
\end{enumerate}
\subsubsection{Selecting $\lambda$}
\label{sssec:selectreg}
The logistic regression models in this study are estimated with an $\ell_2$ (Ridge) regularization penalty to mitigate overfitting from high-dimensional predictors while handling multicollinearity. The associated objective function is the penalized log-likelihood:
\[
\hat\beta(\lambda) = \underset{\beta}{\operatorname{arg max}}\left \{\sum_{t=1}^{T} \left [y_tX^{'}_{t}\beta - \log(1+e^{X^{'}_{t}\beta}) \right ] - \lambda \sum_{j=1}^{N}\beta^2_j \right \},
\]
The hyperparameter $\lambda \geq 0$ controls the strength of the penalty. Due to implementation in scikit-learn, we tune for the parameter $C$, which is equivalent to $\dfrac{1}{\lambda}$. To select $C$ in a manner that is robust and free of look-ahead bias, we perform a time-series cross-validation procedure exclusively on the initial training sample (data from $t=1,..,T$). The selected value, $C^*$, is then held constant throughout the entire recursive out-of-sample forecasting exercise.

The selection procedure is as follows:
\begin{enumerate}
    \item We specify a logarithmic grid of 30 candidate values for the hyperparameter, denoted by the set $\Lambda$, spanning the interval $[10^{-3}, 10^{1}]$.

    \item We use an expanding-window cross-validation scheme with $K=5$ splits, consistent with scikit-learn's \texttt{TimeSeriesSplit}. The initial training sample is partitioned into $K+1=6$ contiguous blocks. For each split $k \in \{1, ..., K\}$, the training set uses the first $k$ blocks and the validation set uses the $(k+1)$-th block.
    
    The size of each validation block, $s$, is approximately $s = \dfrac{T}{K+1} \approx 60$ months. The sets are constructed as:
    \begin{align*}
        S_k^{\text{train}} &= \{t \mid 1 \le t \le s \cdot k \} \\
        S_k^{\text{val}} &= \{t \mid s \cdot k < t \le s \cdot (k+1) \},
    \end{align*}
    ensuring that training data always precedes validation data.

    \item For each candidate value $C_m \in \Lambda$ and for each split $k$, we perform the following steps:
    \begin{enumerate}
        \item Estimate the coefficient vector $\hat{\beta}_k(\lambda_m)$ by maximizing the penalized log-likelihood on the training data $S_k^{\text{train}}$.
        
        \item Generate out-of-sample probability forecasts, $\hat{p}_t(\lambda_m)$, for each observation in the validation set, $t \in S_k^{\text{val}}$.
        
        \item Calculate the Brier Score on the validation set as a measure of forecast performance:
        \[
            \text{BS}_k(C_m) = \frac{1}{|S_k^{\text{val}}|} \sum_{t=1}^{S_k^{\text{val}}} (\hat{p}_t(C_m) - y_t)^2,
        \]
        where $|S_k^{\text{val}}|$ is the number of observations in the validation set.
    \end{enumerate}

    \item The final value $\lambda^*$ is chosen as the candidate that minimizes the average Brier Score across all $K$ folds.
    \[
        C^* = \underset{C_m \in \Lambda}{\arg\min} \left( \frac{1}{K} \sum_{k=1}^{K} \text{BS}_k(C_m) \right)
    \]
\end{enumerate}
This process is then repeated for every predictor set trained on a Logit-$\ell_2$ classifier.

\section{Additional Metrics}

\subsection{Bootstrap confidence interval}

\autoref{tab_app:zt_vs_xt} is the same table as \autoref{tab:zt_vs_xt}, but with confidence intervals estimated based on bootstrapping. 

We produce these confidence intervals for the point estimates of PR AUC and Brier Score using a stationary bootstrap method with 1,000 replications\footnote{In practice, when $L$ is small, the stationary bootstrap occasionally discards a single replication due to edge effects, resulting in 999 rather than 1,000 usable replications. This has no substantive impact on the results.}. We set the average block length as 
\[
L = \max\!\big(h,\; \lfloor T^{1/3} \rceil \big),
\]
where $h$ is the forecast horizon and $T=420$ (our out-of-sample size). The term $T^{1/3}$ follows the heuristic from \citet{politis1994stationary}, which shows that the stationary bootstrap achieves good asymptotic properties when the block length grows at the rate of the sample size raised to the one-third power. However, we set the block length to be at least as large as the horizon $h$ to ensure that each bootstrap sample preserves the dependence structure needed for $h$-step-ahead predictions. \autoref{tab_app:zt_vs_xt} also includes the percentage of bootstrap samples where the benchmark outperformed the proposed model in parentheses.


\begin{table}[t]
\centering
\caption{Out-of-Sample Performance of the At-Risk Transformation ($Z_t$) vs. Benchmarks}
\label{tab_app:zt_vs_xt}
\resizebox{\textwidth}{!}{%
\begin{tabular}{l ccc ccc}
\toprule
& \multicolumn{3}{c}{\textbf{PR AUC}} & \multicolumn{3}{c}{\textbf{Brier Score}} \\
\cmidrule(lr){2-4} \cmidrule(lr){5-7}
\textbf{Model Configuration} & \textbf{$h=3$} & \textbf{$h=6$} & \textbf{$h=12$} & \textbf{$h=3$} & \textbf{$h=6$} & \textbf{$h=12$} \\
\midrule

\textbf{Proposed ($Z_t$, Logit-L2)}& \makecell{\textbf{0.718} \\ \footnotesize{[0.406, 0.897]}} & \makecell{0.370 \\ \footnotesize{[0.154, 0.641]}} & \makecell{\textbf{0.398} \\ \footnotesize{[0.107, 0.647]}} & \makecell{\textbf{0.049} \\ \footnotesize{[0.025, 0.076]}} & \makecell{\textbf{0.082} \\ \footnotesize{[0.044, 0.124]}} & \makecell{\textbf{0.087} \\ \footnotesize{[0.055, 0.121]}} \\
\addlinespace 

Benchmark ($X_t$, Logit-L2) & \makecell{0.501 (2.9\%) \\ \footnotesize{[0.239, 0.775]}}& \makecell{0.286 (16.3\%) \\ \footnotesize{[0.129, 0.526]}} & \makecell{0.170 (2.6\%) \\ \footnotesize{[0.069, 0.336]}} & \makecell{0.069 (0.3\%) \\ \footnotesize{[0.041, 0.099]}} & \makecell{0.096 (12.3\%) \\ \footnotesize{[0.063, 0.131]}} & \makecell{0.121 (0.1\%) \\ \footnotesize{[0.080, 0.161]}} \\

Benchmark (PCA of $X_t$, Logit-L2) & \makecell{0.552 (12.5\%) \\ \footnotesize{[0.273, 0.829]}} & \makecell{\textbf{0.408} (68.4\%) \\ \footnotesize{[0.196, 0.651]}} & \makecell{0.314 (24.8\%) \\ \footnotesize{[0.124, 0.519]}} & \makecell{0.064 (5.1\%) \\ \footnotesize{[0.038, 0.094]}} & \makecell{0.083 (48.8\%) \\ \footnotesize{[0.054, 0.113]}} & \makecell{0.108 (1.3\%) \\ \footnotesize{[0.080, 0.136]}} \\

Benchmark ($X_t$, XGBoost) & \makecell{0.584 (2.4\%) \\ \footnotesize{[0.299, 0.806]}} & \makecell{0.338 (35.6\%) \\ \footnotesize{[0.131, 0.647]}} & \makecell{0.351 (32.9\%)\\ \footnotesize{[0.115, 0.561]}} & \makecell{0.062 (11.4\%) \\ \footnotesize{[0.029, 0.100]}} & \makecell{0.085 (43.6\%) \\ \footnotesize{[0.037, 0.138]}} & \makecell{0.098 (25.1\%) \\ \footnotesize{[0.048, 0.154]}} \\

\bottomrule
\end{tabular}%
}
\end{table}

\subsection{ROC AUC}
\label{ssec:roc}
We now present the ROC (Receiver Operating Characteristic) AUC scores for the models presented in our preliminary comparison (\autoref{tab:zt_vs_xt}) as a comparison to the use of PR AUC in the main text. To compute ROC AUC, we first convert predicted recession probabilities, $\hat{p}_{t}$, into point forecasts using a decision threshold $\delta \in [0,1]$:
\[
\hat{y}_{t}(\delta) = \mathbf{1}\{\hat{p}_{t} \geq \delta\}.
\]
However, instead of calculating recall and precision, for each $\delta$, we simply find the true positive rate and false positive rate: 
\[
T(\delta) = \frac{TP(\delta)}{TP(\delta) + FN(\delta)}, 
\qquad 
F(\delta) = \frac{FP(\delta)}{FP(\delta) + TN(\delta)},
\]
where $TP$, $TN$, $FN$, and $FP$ denote true positives, true negatives, false negatives, and false positives, and $T$ and $F$ denote the true positive rate and false positive rate, respectively. 
Varying $\delta$ produces the ROC curve. 
The area under the curve is then defined as
\[
\text{ROC AUC} = \int_{0}^{1} T(\delta) \, \frac{dF(\delta)}{d\delta} \, d\delta
= \int T(F) \, dF.
\]
The results are shown in \autoref{tab:zt_vs_xt_roc}. 
\begin{table}[t!]
\centering
\caption{\autoref{tab:zt_vs_xt} ROC AUC Scores}
\label{tab:zt_vs_xt_roc}
\begin{tabular}{l ccc}
\toprule
& \multicolumn{3}{c}{\textbf{ROC AUC}} \\
\cmidrule(lr){2-4} 
\textbf{Model Configuration} & \textbf{$h=3$} & \textbf{$h=6$} & \textbf{$h=12$} \\
\midrule
\textbf{Proposed ($Z_t$, Logit-$\ell_{2}$)} & 
\textbf{0.949} & 
0.861 & 
0.826 \\
\addlinespace
\midrule
\textbf{Alternative Specifications} &&& \\
$X_t$, Logit-$\ell_{2}$ & 
0.913 & 
0.801 & 
0.753 \\
PCA of $X_t$, Logit-$\ell_{2}$ & 
0.923 & 
\textbf{0.895} & 
\textbf{0.843} \\
$X_t$, XGBoost & 
0.920 & 
0.829 & 
0.822 \\
\bottomrule
\end{tabular}
\end{table}
The ranking mostly reflects our preliminary results, with the exception of PCA on $X_t$ performing slightly better on the longer horizons. However, ROC AUC does not consider class imbalance, so models that do well on large expansion periods but perform poorly on the short, sparse recessionary periods can still achieve a high ROC AUC score (part of the reason why we consider PR AUC in the main text).

For the other comparisons, we find that, as in to \autoref{tab:zt_vs_xt_roc}, the results reflect original rankings closely.

\section{Robustness checks}
\subsection{Varying $\tau_{g}$}
\label{ssec:varythresholds}
For robustness, we also evaluate model performance using two other thresholding strategies, computing: (1) cutoff quantiles for each sector and (2) cutoff points for each variable. The results are shown in \autoref{tab:robustness_thresholds}.

\begin{table}[h!]
  \centering
  \caption{Robustness to Alternative Threshold Specifications}
  \label{tab:robustness_thresholds}
  \sisetup{
    table-align-text-post=false,
    table-number-alignment = center
  }
  \begin{tabular}{l *{3}{S[table-format=1.3]} *{3}{S[table-format=1.3]}}
    \toprule
    & \multicolumn{3}{c}{\textbf{PR AUC}} & \multicolumn{3}{c}{\textbf{Brier Score}} \\
    \cmidrule(lr){2-4} \cmidrule(lr){5-7}
    Threshold Specification & {$h=3$} & {$h=6$} & {$h=12$} & {$h=3$} & {$h=6$} & {$h=12$} \\
    \midrule
    \multicolumn{7}{l}{\textbf{Disaggregated $Z_t$ + Logit-$\ell_2$ Model}} \\
    \midrule
    Global Threshold (Main Spec.) & \textbf{0.718} & 0.370 & 0.398 & \textbf{0.049} & 0.082 & 0.087 \\
    Sector-Specific Thresholds & 0.652 & \textbf{0.397} & \textbf{0.508} & 0.052 & \textbf{0.078} & \textbf{0.085} \\
    Variable-Specific Thresholds & 0.662 & 0.370 & 0.313 & 0.053 & \textbf{0.078} & 0.098 \\
    \midrule
    \multicolumn{7}{l}{\textbf{PCA ($Z_t$) + Logit-$\ell_2$ Model}} \\
    \midrule
    Global Threshold (Main Spec.) & 0.688 & 0.528 & 0.404 & 0.062 & 0.081 & 0.106 \\
    Sector-Specific Thresholds & \textbf{0.728} & \textbf{0.572} & \textbf{0.549} & \textbf{0.051} & \textbf{0.068} & \textbf{0.088} \\
    Variable-Specific Thresholds & 0.722 & 0.453 & 0.364 & 0.054 & 0.079 & 0.114 \\
    \bottomrule
  \end{tabular}
\end{table}

The results show that while increasing threshold complexity can help to an extent, it can also increase the chances of overfitting. For both the disaggregated and PCA-based models, using a sector-specific threshold can be of benefit, particularly at longer horizons. However, implementing a variable-specific cutoff leads to overfitting, as the threshold for each variable in the initial training sample tends to evolve over time. Other specialized thresholding techniques could also be explored in future research.

\subsection{Varying $h_{g}$}
\label{ssec:varysmooth}
In Section \ref{ssec:params}, we briefly discussed the $h_g$ tuning parameter, which denotes the moving average window of the input signal. In our main analysis, we set $h_g=1$ and added explicit lags; now, we explore an alternative way to incorporate past information without explicit lags by setting $h_g>1$, shown in \autoref{tab:robustness_ma}. 

The results demonstrate that while using a moving average ($h_g > 1$) can incorporate past information, it generally leads to lower PR AUC and higher Brier Scores compared with our main specification ($h_g = 1$ with explicit lags on $Z_t$). This suggests that for the ``at-risk'' 'transformation, the predictive signal derived from the timing and persistence of the binarized `'at-risk'' states (captured by lags of $Z_t$) is more valuable than smoothing the intensity of the continuous variable itself before binarization.

\begin{table}[t!]
  \centering
  \caption{$h_g$ Moving Average Results}
  \label{tab:robustness_ma}
  \sisetup{
    table-align-text-post=false,
    table-number-alignment = center
  }
  \begin{tabular}{l *{3}{S[table-format=1.3]} *{3}{S[table-format=1.3]}}
    \toprule
    & \multicolumn{3}{c}{\textbf{PR AUC}} & \multicolumn{3}{c}{\textbf{Brier Score}} \\
    \cmidrule(lr){2-4} \cmidrule(lr){5-7}
    Moving Average Window & {$h=3$} & {$h=6$} & {$h=12$} & {$h=3$} & {$h=6$} & {$h=12$} \\
    \midrule
    \multicolumn{7}{l}{\textbf{Disaggregated $Z_t$ + Logit-$\ell_2$ Model}} \\
    \midrule
    $h_g=3$ & 0.551 & \textbf{0.370} & 0.363 & \textbf{0.066} & 0.102 & 0.142 \\
    $h_g=6$ & \textbf{0.611} & 0.257 & \textbf{0.479} & 0.067 & 0.110 & \textbf{0.124} \\
    $h_g=12$ & 0.563 & 0.330 & 0.254 & 0.074 & \textbf{0.097} & 0.135 \\
    \midrule
    \multicolumn{7}{l}{\textbf{PCA ($Z_t$) + Logit-$\ell_2$ Model}} \\
    \midrule
   $h_g=3$ & \textbf{0.684} & 0.411 & 0.283 & \textbf{0.086} & 0.121 & 0.161 \\
   $h_g=6$ & 0.618 & \textbf{0.481} & \textbf{0.288} & 0.090 & \textbf{0.113} & \textbf{0.149} \\
   $h_g=12$ & 0.578 & 0.254 & 0.167 & 0.099 & 0.134 & 0.168 \\
    \bottomrule
  \end{tabular}
\end{table}

\subsection{Contemporaneous Predictors}
In our main analysis, all models in the study were evaluated with lags of $L=\{3,6,12\}$. As a robustness check, we also evaluate the models without any added lags. In \autoref{tab:zt_vs_xt_unlagged}, we present our baseline comparisons using only contemporaneous signals.
\begin{table}[h!]
\centering
\caption{Baseline Performance Without Lags}
\label{tab:zt_vs_xt_unlagged}
\begin{tabular}{l ccc ccc}
\toprule
& \multicolumn{3}{c}{\textbf{PR AUC}} & \multicolumn{3}{c}{\textbf{Brier Score}} \\
\cmidrule(lr){2-4} \cmidrule(lr){5-7}
\textbf{Model Configuration} & \textbf{$h=3$} & \textbf{$h=6$} & \textbf{$h=12$} & \textbf{$h=3$} & \textbf{$h=6$} & \textbf{$h=12$} \\
\midrule

\textbf{Proposed ($Z_t$, Logit-$\ell_{2}$)} & 
\textbf{0.594} & 
\textbf{0.464} & 
0.274 & 
\textbf{0.069} & 
\textbf{0.102} & 
\textbf{0.137} \\
\addlinespace

\midrule
\textbf{Alternative Specifications} &&&&& \\
$X_t$, Logit-$\ell_{2}$ & 
0.462 & 
0.332 & 
0.311 & 
0.087 & 
0.107 & 
0.162 \\

PCA of $X_t$, Logit-$\ell_{2}$ & 
0.541 & 
0.429 & 
\textbf{0.380} & 
0.105 & 
0.110 & 
0.146 \\

$X_t$, XGBoost & 
0.582 & 
0.275 & 
0.212 & 
0.066 & 
0.104 & 
0.120 \\

\bottomrule
\end{tabular}
\footnotesize
\begin{flushleft}
\textit{Note:} This table compares the out-of-sample performance of a Logistic Regression with $\ell_{2}$ penalty trained on our binary at-risk state matrix ($Z_t$) against several benchmarks (using unlagged predictors). Bold values indicate the best-performing model for each metric and horizon. 
\end{flushleft}
\end{table}
In most places, there is a drop in performance, but with some models on certain horizons (such as $Z_t$ on $h=6$) the predictive performance increases. However, we see most models generally perform better when lagged features are included. Below is \autoref{tab:aggregation_unlagged} (aggregation strategies) using only contemporaneous features.

\begin{table}[h!]
  \centering
  \caption{Aggregation Strategies for $Z_t$ Without Lags}
  \label{tab:aggregation_unlagged}
  \resizebox{\textwidth}{!}{
  \begin{tabular}{l l ccc c ccc}
    \toprule
    & & \multicolumn{3}{c}{\textbf{PR AUC}} & & \multicolumn{3}{c}{\textbf{Brier Score}} \\
    \cmidrule(r){3-5} \cmidrule(l){7-9}
    \textbf{Aggregation Method} & \textbf{Model} & \textit{$h=3$} & \textit{$h=6$} & \textit{$h=12$} & & \textit{$h=3$} & \textit{$h=6$} & \textit{$h=12$} \\
    \midrule
    Disaggregated ($Z_t$) & Logit-$\ell_2$ & 0.594 & 0.464 & 0.274 & & 0.069 & 0.102 & 0.137 \\
                                & XGBoost & 0.659 & 0.281 & 0.239 & & \textbf{0.055} & \textbf{0.089} & 0.106 \\
    \addlinespace
    Simple Average & Logit & 0.555 & 0.362 & 0.104 & & 0.154 & 0.205 & 0.253 \\
                     & XGBoost & 0.322 & 0.276 & 0.088 & & 0.191 & 0.207 & 0.251 \\
    \addlinespace
    PCA on $Z_t$ & Logit-$\ell_2$ & \textbf{0.668} & \textbf{0.507} & 0.289 & & 0.083 & 0.123 & 0.153 \\
                       & XGBoost & 0.583 & 0.344 & \textbf{0.312} & & 0.068 & 0.091 & \textbf{0.091} \\
    \bottomrule
  \end{tabular}
  }
  \parbox{\textwidth}{\footnotesize \textit{Notes:} This table compares the aggregation and modeling strategies applied to the $Z_t$ matrix without lags. No penalization is used for the simple average, since there is only one feature.}
\end{table}
The discriminatory power significantly decreases for XGBoost, which relies on the lags as mentioned above, but feature sets using it have the lowest Brier Scores at each horizon. As mentioned previously, although removing lags improves performance in certain scenarios, a general conclusion is that all models perform better when incorporating lagged features, and $Z_t$ still offers better predictive performance than the benchmarks in most cases, even without lags.

\subsection{Other Parsimonious Sets}
\label{ssec:otherparsimony}
We demonstrate the use of only one representative, parsimonious set in Section \ref{sec:parsimony}, but we test alternative subsets to see whether $Z_t$ surpasses $X_t$ there as well.

For set $S_t$, we only use spreads\footnote{We use five yield spreads (10-year, 5-year, 1-year, 6-month, and 3-month Treasury-Fed Funds spreads), two credit spreads (AAA- and Baa-Fed Funds spreads), and the Commercial Paper-Fed Funds spread.}. For set $R_t$, we use only real economy variables\footnote{We use four labor market (payrolls, unemployment insurance claims, average weekly hours in manufacturing, and the unemployment rate), two output (industrial production, real personal income), housing starts, and retail sales.}. \autoref{tab:subset_extra} shows the key metrics for our competing specifications.

\begin{table}[h!]
  \centering
  \caption{Out-of-Sample Performance of Parsimonious Models with Alternative Indicators}
  \label{tab:subset_extra}
  \sisetup{
    table-align-text-post=false,
    table-number-alignment = center
  }
  \begin{tabular}{l *{3}{S[table-format=1.3]} *{3}{S[table-format=1.3]}}
    \toprule
    & \multicolumn{3}{c}{\textbf{PR AUC}} & \multicolumn{3}{c}{\textbf{Brier Score}} \\
    \cmidrule(lr){2-4} \cmidrule(lr){5-7}
    Model Specification & {$h=3$} & {$h=6$} & {$h=12$} & {$h=3$} & {$h=6$} & {$h=12$} \\
    \midrule
    \multicolumn{7}{l}{\textit{Full Feature Space}} \\
    \quad Continuous ($X_t$) + XGBoost & 0.584 & 0.338 & 0.351 & 0.062 & 0.085 & 0.098 \\
    \quad ``At-Risk'' ($Z_t$) + Logit-$\ell_{2}$ & \textbf{0.718} & \textbf{0.370} & \textbf{0.398} & \textbf{0.049} & \textbf{0.082} & \textbf{0.087} \\
    \midrule
    \multicolumn{7}{l}{\textit{$S_t$ (8 Spreads)}} \\
    \quad Continuous ($X_t$) + XGBoost  & 0.490 & 0.356 & 0.416 & 0.079 & \textbf{0.085} & \textbf{0.086} \\
    \quad ``At-Risk'' ($Z_t$) + Logit-$\ell_2$  & \textbf{0.529} & \textbf{0.427} & \textbf{0.451} & \textbf{0.071} & 0.086 & 0.103 \\
    \midrule
    \multicolumn{7}{l}{\textit{$R_t$ (8 Real Economy Indicators)}} \\
    \quad Continuous ($X_t$) + XGBoost & 0.439 & 0.208 & \textbf{0.131} & \textbf{0.077} & \textbf{0.123} & \textbf{0.103} \\
    \quad ``At-Risk'' ($Z_t$) + Logit-$\ell_{2}$ & \textbf{0.542} & \textbf{0.238} & 0.107 & 0.128 & 0.183 & 0.233 \\
    \bottomrule
    \addlinespace
  \end{tabular}
    \parbox{\textwidth}{\footnotesize 
  \textit{Notes:} The ``Full Feature Space'' corresponds to the entire FRED-MD panel used in earlier sections. 
  The $S_t$ models use five yield spreads (10-year, 5-year, 1-year, 6-month, and 3-month Treasury-Fed Funds spreads), two credit spreads (AAA- and Baa-Fed Funds spreads), and the Commercial Paper-Fed Funds spread. The $R_t$ models use four labor market indicators (payrolls, unemployment insurance claims, average weekly hours in manufacturing, and the unemployment rate), two output indicators (industrial production, real personal income), housing starts, and retail sales.
  }
\end{table} 

For $S_t$, while performance decreases on the short-term horizons for all models (\citet{estrella1996yield} find that yield curve performance is higher after 2 quarters), we find that the longer-horizon models do well. Even using just the spreads, the ``at-risk'' transformation delivers superior discriminatory power on $h=3,6,12$. Its probability estimates are slightly less accurate on $h=12$, but this is compensated for through its better ability to separate the two classes.

For set $R_t$, the performance is slightly more variable, with the ``at-risk'' set showing superior discriminatory power on shorter horizons, while the continuous version of $R_t$ exhibits better calibration at the significantly high cost of discriminatory power. While it may seem like the continuous $R_t$ is better at longer horizons, performance is so low that it is not realistic to compare them—it is not even realistic to use them in a model for recession prediction at these ranges given their empirical performance. When considering a representative set that performs well and is part of a more realistic recession forecasting exercise, we find that the ``at-risk'' transformation is beneficial for model performance.

\section{Additional Survey of Predictor Representation} 
\label{ssec:additional_lit}
This appendix presents a survey of the standard treatment of predictors in existing literature on model-based recession forecasting. We specifically focused on the \textit{International Journal of Forecasting}, a leading journal in the field. Our search included all articles as of September 2025 containing the terms ``recession'' and ``probit''. This was supplemented by the first 100 relevant Google Scholar results (using the query: \texttt{source:"international journal of forecasting" "recession"}, without the term ``probit'').
The final sample consists of 24 papers, and they are listed below. These papers can be grouped into three broad topics: (i) variable selection for recession forecasting using probit/logit regression, (ii) applications of new statistical or machine learning methods to binary recession outcomes, and (iii) real-time recession dating (nowcasting) using Markov-switching models. 

Whether the prediction model is linear or nonlinear, we find that none of these papers apply explicit nonlinear transformations of predictors beyond standardization and stationarization, in contrast to the approach proposed in our paper.

\begin{enumerate}
\item \cite{liu2016}; Probit models map a large panel of financial and macro indicators to the NBER recession state; US monthly 1959–2011, predictors stationarized per series (levels, log‐diffs, annual diffs, moving averages); target is NBER recession dummy.
\item \cite{DuarteVenetisPaya2005}; Dynamic probit (and threshold growth models) use the EMU term spread to predict recessions constructed from GDP contractions; Euro area quarterly 1970Q1–2000Q4; recession dummy = $\ge$2 negative quarters in 5‐quarter window; GDP in log‐diffs (annualized), spreads in levels.
\item \cite{Hamilton2011}; Markov‐switching models of real activity detect recession states in real time from coincident indicators; US quarterly/monthly real activity, unemployment, term structure; growth via log‐diffs/annualization; nonlinearity via regime switching.
\item \cite{Dopke2017}; Boosted regression trees classify the binary recession state and capture nonlinear thresholds/interactions; Germany monthly 1973–2014; 35 indicators, many YoY or real terms, some in levels; stationarity checked; nonlinearity from regression trees.
\item \cite{AastveitAnundsenHerstad2019}; Probit tests whether residential investment improves recession prediction over standard predictors; 12 OECD countries 1960Q1–2014Q4; NBER/ECRI dummies; predictors in quarterly log differences with lags up to 4.
\item \cite{Qi2001}; A feedforward neural network combines leading indicators (esp. yield spread) to generate multi‐quarter recession probabilities; US quarterly 1967Q2–1995Q1; 27 leading indicators, many log‐diffs or growth, some levels; NN provides nonlinearity.
\item \cite{AastveitJoreRavazzolo2016}; Markov‐switching factor models identify Norwegian business‐cycle regimes and turning points; Norway quarterly 1978–2011; GDP, macro, surveys, FCI; logged/factored data; regime switching supplies nonlinearity.
\item \cite{BluedornDecressinTerrones2016}; Logit predicts recession starts using asset price drops, volatility, spreads, oil prices; G7 quarterly 1970–2011; asset prices deflated, oil real, spreads in levels; logit nonlinearity only.
\item \cite{LiShengYang2021}; Bayesian quickest detection (sequential) rule on HMM flags peaks/troughs earlier than DFMS; US monthly vintages 1967–2013; four coincident series; Kalman filtering + factor extraction; no extra nonlinear transforms.
\item \cite{DavigHall2019}; Naïve Bayes and MS‐Naïve Bayes classify recessions with macro‐financial predictors, exploiting persistence; US monthly 1959–2016; FRED‐MD standardized transforms; 10 lags; nonlinear via NB/MS.
\item \cite{FintzenStekler1999}; Decision‐theoretic and survey‐based analysis explains why forecasters missed the 1990 recession; US 1989–1990 survey + Greenbook; macro in growth rates; descriptive, no model transforms.
\item \cite{Levanon2015}; A PCA‐based Leading Credit Index replaces M2 in the LEI, probits on the new LEI yield earlier recession signals; US 1990–2013 monthly; six standardized financial indicators interpolated; probit on lagged LEI.
\item \cite{HuangStartz2020}; MS factor model augmented with regime‐switching volatility improves real‐time dating; US monthly 1959–2018; four coincident series in log‐diffs, stock returns via MS volatility.
\item \cite{Carstensen2020}; Three‐state MS‐DFM distinguishes ordinary vs severe recessions; Germany monthly 1991–2016; 35 candidates, EN picks 6; hard series in growth, spreads in levels; standardized.
\item \cite{ClementsHarvey2011}; Combining logit‐based probabilities (spread \& hours) via log‐odds improves forecast scores; US monthly 1965–2007; spread in levels, hours in YoY\%; nonlinearity from logit link and KK combination.
\item \cite{GiustoPiger2017}; Learning Vector Quantization classifies monthly states from coincident growth rates with persistence rule; US monthly 1967–2013; four coincident indicators in growth; vintages mimic real time.
\item \cite{ProanoTheobald2014}; Composite real‐time dynamic probit pools multiple lag structures/indicators for stable probabilities; Germany 1991–2011 \& US 1969–2011 monthly; industrial production/NBER dummies; growth rates, spreads in levels, standardized.
\item \cite{Hansen2024}; Static/dynamic probits use VIX–yield spread cycles to outperform the spread alone; US monthly 1950–2022; 10y–3m spread, VIX (extended pre‐1990); smoothed, standardized, winsorized cycle indicators.
\item \cite{Antunes2018}; Dynamic panel probits with lags + exuberance dummies improve banking‐crisis early warnings; 22 European countries quarterly 1970–2012; credit gaps, debt service, house prices, equities; ratios, YoYs, HP‐filtered gaps.
\item \cite{ChauvetPotter2010}; Probits with recurrent breaks (and AR latent) improve classification over fixed/no‐break models; US monthly 1959–2007; four coincident series in 100×log‐diffs; break‐specific variances, AR latent adds persistence.
\item \cite{LaytonKatsuura2001}; Regime‐switching with time‐varying transition probabilities outperforms probit/logit for dating/forecasting; US monthly 1949–1999; ECRI coincident/leading composites; month‐to‐month growth, moving sums for LRG.
\item \cite{CarrieroMarcellino2007}; Probits on selected indicators outperform VARs/MS for UK turning points; UK monthly 1978–2004; coincident (IP, sales, employment, income) + leading (CB set); growth/log‐diffs, standardized.
\item \cite{LahiriYang2015}; Probits/MS with LCI‐augmented LEI yield better probabilities (via QPS/ROC diagnostics); US monthly 1990–2014; LEI in log‐diffs; focus on resolution/discrimination rather than predictor transforms.
\item \cite{DePaceWeber2016}; Probits and TVP models show HY spread predictive power faded post‐2000 while term spread regained long horizon signal; US quarterly GDP 1982–2011 \& monthly IP 1982–2011; HY/term spreads stationary, growth annualized.
\end{enumerate}

\end{document}